\documentclass[oneside,english]{elsart}
\usepackage[T1]{fontenc}
\usepackage[latin1]{inputenc}
\usepackage{subfigure}
\usepackage{amsmath}
\usepackage{graphicx}
\usepackage{amssymb}
\usepackage[authoryear]{natbib}

\makeatletter

\providecommand{\tabularnewline}{\\}

\usepackage{mdwlist}

\usepackage{babel}
\makeatother
\begin{document}
\begin{frontmatter}

\title{Effect of DNA looping on the induction kinetics of the \emph{lac}
operon}

\author{Atul Narang}

\address{Department of Chemical Engineering, University of Florida, Gainesville,
FL~32611-6005.}

\ead{narang@che.ufl.edu}

\ead[url]{http://narang.che.ufl.edu}

\thanks{Corresponding author. Tel: + 1-352-392-0028; fax: + 1-352-392-9513}

\begin{keyword}
\noindent Mathematical model, bacterial gene regulation, transcriptional
regulation, induction, lac operon, DNA looping
\end{keyword}
\begin{abstract}
The induction of the \emph{lac} operon follows cooperative kinetics.
The first mechanistic model of these kinetics is the \emph{de facto}
standard in the modeling literature (Yagil \& Yagil, Biophys J, 11,
11-27, 1971). Yet, subsequent studies have shown that the model is
based on incorrect assumptions. Specifically, the repressor is a
tetramer with four (not two) inducer-binding sites, and the operon
contains two auxiliary operators (in addition to the main operator).
Furthermore, these structural features are crucial for the formation
of DNA loops, the key determinants of \emph{lac} repression and
induction. Indeed, the repression is determined almost entirely
(>95\%) by the looped complexes (Oehler et al, EMBO J, 13, 3348,
1990), and the pronounced cooperativity of the induction curve
hinges upon the existence of the looped complexes (Oehler et al,
Nucleic Acids Res, 34, 606, 2006). Here, we formulate a model of
\emph{lac} induction taking due account of the tetrameric structure
of the repressor and the existence of looped complexes. We show
that: (1) The kinetics are significantly more cooperative than those
predicted by the Yagil \& Yagil model. The cooperativity is higher
because the formation of looped complexes is easily abolished by
repressor-inducer binding. (2) The model provides good fits to the
repression data for cells containing tetrameric (or mutant dimeric)
repressor, as well as the induction curves for 6 different strains
of \emph{E. coli}. It also implies that the ratios of certain looped
and non-looped complexes are independent of inducer and repressor
levels, a conclusion that can be rigorously tested by gel
electrophoresis. (3) Repressor overexpression dramatically increases
the cooperativity of the induction curve. This suggests that
repressor overexpression can induce bistability in systems, such as
growth of \emph{E. coli} on lactose, that are otherwise monostable.
\end{abstract}
\end{frontmatter}

\section{Introduction\label{s:Intro}}

Genetic switches plays a fundamental role in development and evolution~\citep{Carroll,Ptashne2}.
The development of embryos is now known to be orchestrated by an array
of genetic switches. There is growing belief that the biodiversity
of organisms reflects the evolution of the regulatory genes controlling
these genetic switches.

The \emph{lac} operon is a paradigm of the mechanism by which genetic
switches are regulated. Key mechanisms of gene regulation, such as
negative and positive control by the repressor and CAP, respectively,
were revealed by studies of the \emph{lac} operon~\citep{Muller-Hill}.
Not surprisingly, the \emph{lac} operon has been, and continues to
be, the system of choice for researchers interested in the dynamics
of gene regulation~\citep{Laurent2005}.

\begin{figure}
\noindent \begin{centering}\includegraphics[width=10cm,height=1.5cm]{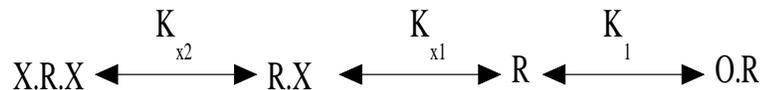}\par\end{centering}

\caption{\label{f:SchemeYagil}Kinetic scheme of the Yagil \& Yagil model~\citep{yagil71}.
Here, $X$ denotes the inducer, $R$ denotes the repressor, and $O$
denotes the operator. }
\end{figure}

It has been known for many years that the \emph{lac} induction rate
is a sigmoidal function of the inducer concentration~\citep{Herzenberg1959}.
The first mechanistic model of these kinetics was based on the following
assumptions (Fig.~\ref{f:SchemeYagil}):

\begin{enumerate}
\item The \emph{lac} operon contains one operator.
\item The \emph{lac} repressor has two inducer-binding sites.
\item Inducer-bound repressor ($R\cdot X$, $X\cdot R\cdot X$) cannot bind
to the operator.
\end{enumerate}
The first assumption was supported by the prevailing knowledge of
the \emph{lac} operon. There was no direct evidence for the last two
assumptions --- they were made because they yielded sigmoidal kinetics.
Indeed, the above assumptions imply that the induction rate is proportional
to the expression \begin{equation}
\frac{1+K_{x1}x+K_{x1}K_{x2}x^{2}}{1+K_{1}r_{t}+K_{x1}x+K_{x1}K_{x2}x^{2}}\label{eq:YagilGeneral}\end{equation}
where $x$ is the inducer concentration; $K_{x1},K_{x2}$ are the
association constants for binding of the first and the second inducer
molecules to the repressor; $K_{1}$ is the association constant for
repressor-operator binding; and $r_{t}$ is the total concentration
of the repressor.

Yagil \& Yagil also performed an extensive study of the extent to
which their model captured the data. They showed that in some instances,
the data could be fitted by the simpler expression\begin{equation}
\frac{1+K_{x1}K_{x2}x^{2}}{1+K_{1}r_{t}+K_{x1}K_{x2}x^{2}}\label{eq:YagilSpecial}\end{equation}
which does not contain the linear term, $K_{x1}x$. In yet other cases,
the data could not be fitted unless eq.~(\ref{eq:YagilGeneral})
was used. Nevertheless, eq.~(\ref{eq:YagilSpecial}) has become the
\emph{de facto} standard in the modeling literature~\citep{chung96,ozbudak04}.

Since the publication of Yagil \& Yagil's paper, studies have shown
that assumptions (1)--(3) of the Yagil \& Yagil model are not consistent
with the structure of the \emph{lac} operator and repressor. Specifically,
the \emph{lac} operon contains not one, but three operators; the repressor
contains not two, but four inducer-binding sites; and finally, inducer-bound
repressor can bind to the operator. Furthermore, these structural
features have a profound effect on the repression and induction of
the \emph{lac} operon.

\begin{figure}
\noindent \begin{centering}\includegraphics[width=3.5in,keepaspectratio]{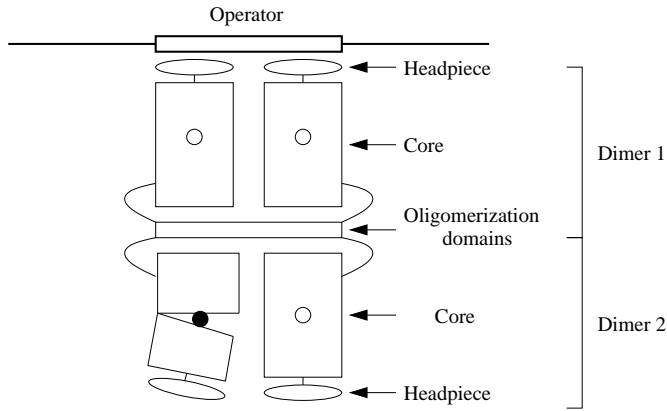}\par\end{centering}

\caption{\label{fig:StructureOfRepressor}The structure of the \emph{lac}
repressor (adapted from~\citealp[Chap.~3.4]{Muller-Hill}). The open
circles represent free inducer-binding sites. The binding of an inducer
to a dimer (closed circle) changes the relative orientation of the
two subdomains of the core, thus separating the headpieces and abolishing
their ability to bind to an operator.}
\end{figure}

\emph{In vivo}, the \emph{lac} repressor is a tetrameric molecule~\citep{Barry1999},
which can be viewed as a {}``dimer of dimers'' (Fig.~\ref{fig:StructureOfRepressor}).
Each monomer contains a \emph{headpiece} that can bind to the operator,
a \emph{core} containing an inducer-binding site, and an \emph{oligomerization
domain} that mediates the linking of the two dimers. If a repressor
dimer is inducer-free, its headpieces can interact strongly with an
operator. This interaction is reduced if the dimer is inducer-bound,
because inducer binding changes the relative orientation of the two
subdomains of the core, thus increasing the distance between the headpieces
of the dimer~\citep[Fig.~17]{Lewis2005}. Kinetic studies suggest
that the binding of even one inducer molecule to a dimer abolishes
its ability to bind to an operator~\citep{Oehler2006}. It is therefore
clear that the repressor molecule has 4 identical inducer binding
sites, and inducer-bound repressor can bind to the operator, provided
one of its dimers is inducer-free.

\begin{figure}
\noindent \begin{centering}\includegraphics[width=3.5in,keepaspectratio]{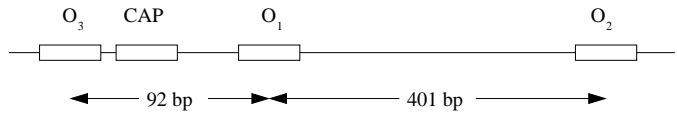}\par\end{centering}

\caption{\label{f:OperatorLoci}The arrangement of the \emph{lac} operators
(not drawn to scale). The main operator, $O_{1}$, lies within the
\emph{lac} promoter. The auxiliary operator, $O_{2}$, lies within
\emph{lacZ}, the gene encoding $\beta$-galactosidase, and the auxiliary
operator, $O_{3}$ is adjacent to the binding site for CAP.}
\end{figure}

It has also been found that in addition to the \emph{main} operator,
denoted $O_{1}$, there are two \emph{auxiliary} operators, denoted
$O_{2}$ and $O_{3}$ (Fig.~\ref{f:OperatorLoci}). The auxiliary
operator, $O_{2}$, located 401~bp downstream of $O_{1}$, lies within
\emph{lacZ}, the gene encoding $\beta$-galactosidase. The auxiliary
operator, $O_{3}$, located 92~bp upstream of $O_{1}$, is adjacent
to the CAP binding site. Given these locations, one expects the transcriptional
repression to increase in the presence of the auxiliary operators.
If the repressor binds to $O_{2}$, it can hinder the transcription
of the operon; if it binds to $O_{3}$, CAP cannot attach effectively
to its cognate site. It turns out that the repression is indeed higher
in the presence of the auxiliary operators, but not because these
operators have a strong affinity for the repressor. Instead, they
increase the repression by a subtle interaction that stabilizes the
binding of the repressor to $O_{1}$.

This interaction was revealed by measuring the \emph{repression} in
cells containing various combinations of operators~\citep{Oehler1990}.
The repression is defined as the ratio \begin{equation}
\mathcal{R}\equiv\frac{\left.e\right|_{x\rightarrow\infty}}{\left.e\right|_{x=0}}\label{eq:Rdefn}\end{equation}
where $x$ is the concentration of a gratuitous inducer (IPTG in these
experiments), and $e$ is the specific $\beta$-galactosidase activity
measured during exponential growth of \emph{lacY}$^{-}$ cells on
a mixture of IPTG and a carbon source that cannot induce \emph{lac}
transcription (glycerol in these experiments). It provides a measure
of the transcriptional inhibition in the absence of the inducer: $\mathcal{R}$
is~1 if there is no inhibition, and becomes progressively higher
with the strength of the inhibition. Oehler et al observed that (Table~\ref{t:RepressionData}):

\begin{enumerate}
\item In the absence of the auxiliary operators, the repression is only
18. However, it increases dramatically if $O_{2}$ or $O_{3}$ are
also present ($\sim$40- and $\sim$25-fold, respectively).
\item In the presence of only $O_{2}$ or $O_{3}$, the repression is similar
that observed in cells lacking all three operators. Thus, $O_{2}$
and $O_{3}$ have almost no affinity for the repressor.\\
It follows that the increased repression observed in the presence
of $O_{1}$ and $O_{2}$ (or $O_{3}$) does not occur simply because
the auxiliary operators have a strong affinity for the repressor ---
instead, there is some interaction between the operators.
\item The repression in the presence of $O_{2}$ and $O_{3}$ is also similar
to basal levels. It follows that the interaction primarily involves
the pairs, $O_{1},O_{2}$ and $O_{1},O_{3}$ --- interactions between
$O_{2}$ and $O_{3}$ make almost no contribution to the repression.
\item In the presence of all three operators, the repression is only 2-
or 3-fold higher than that observed in the presence of the pairs,
$O_{1},O_{2}$ and $O_{1},O_{3}$. Thus, the presence of either one
of these two pairs is sufficient for the bulk of the repression.
\end{enumerate}
Oehler et al argued that the interaction between the operators reflects
the formation of DNA loops.

\begin{table}

\caption{\label{t:RepressionData}Repression observed in the presence of various
combinations of the operators~\citep[Fig.~2]{Oehler1990}.}

\begin{centering}\begin{tabular}{|c|c|c|c|}
\hline
Combination of operators&
Repression&
Combination of operators&
Repression\tabularnewline
\hline
\hline
$O_{1}$&
18&
$O_{3}$&
1\tabularnewline
\hline
$O_{1},$ $O_{2}$&
700&
$O_{2}$, $O_{3}$ &
1.9\tabularnewline
\hline
$O_{1},$ $O_{3}$&
440&
$O_{1}$, $O_{2}$, $O_{3}$&
1300\tabularnewline
\hline
$O_{2}$&
1&
No operators&
1\tabularnewline
\hline
\end{tabular}\par\end{centering}
\end{table}

\begin{figure}
\noindent \begin{centering}\includegraphics[width=3.5in,keepaspectratio]{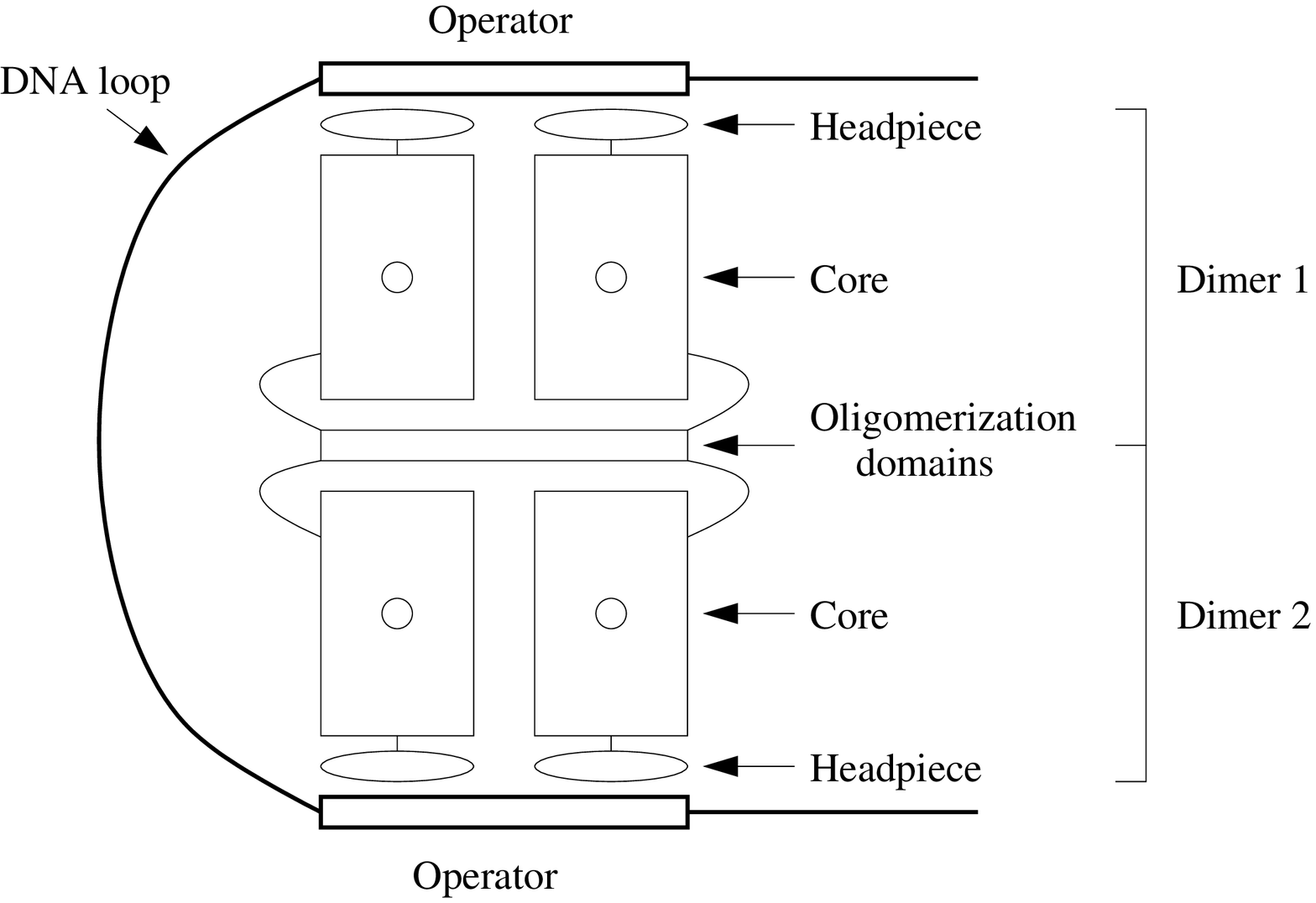}\par\end{centering}

\caption{\label{f:DNAloopFormation}The formation of a DNA loop (from~\citealp[Chap.~3.4]{Muller-Hill}).}
\end{figure}

DNA loops can form only if the repressor is completely free of inducer.
In this case, the binding of one of the repressor dimers to an operator
brings the other (free) dimer close to the remaining the remaining
two operators. If one of these operators is free, the free dimer can
bind to it, thus forcing the intervening DNA to form a loop (Fig.~\ref{f:DNAloopFormation}).

Given the above mechanism for DNA loop formation, Oehler et al explained
their data as follows. The repressor binds primarily to $O_{1}$.
The $O_{1}\cdot R$ complex thus formed is rapidly converted to a
stable DNA loop by interaction with $O_{2}$ or $O_{3}$. The conversion
to a loop is rapid because it is driven by the {}``local concentration''
of $O_{1}\cdot R$ within small spheres having radii equal to the
inter-operator distances of 401 and 92~bp \citep[Fig.~7]{Oehler1994}.
The loop is stable because even if thermal fluctuations cause the
repressor to detach from, say, $O_{1}$, weak interaction of the repressor
with $O_{2}$ or $O_{3}$ keeps it within a small neighborhood of
$O_{1}$, thus increasing the probability that it rebinds to $O_{1}$~\citep[p.~20]{Ptashne2}.
In other words, the local concentration effect increases the {}``on''
rate for loop formation, and the rebinding effect decreases the {}``off''
rate for loop formation. The net result is a high association constant
for loop formation, a fact that is confirmed by the parameter estimates
(Section~\ref{s:Results}).

The above explanation assumes that (a)~despite the low affinity,
the repressor does bind to the auxiliary operators, and (b)~the stability
of the loop rests upon the proximity of the main and auxiliary operators.
Both assumptions were confirmed in subsequent experiments~\citep{Oehler1994}.
It was shown that the repressor binds weakly to the auxiliary operators,
and there is no repression if the auxiliary operators are moved far
away (>3600 base pairs) from $O_{1}$.

Vilar \& Leibler formulated a statistical thermodynamic model to account
for the foregoing repression data~\citep{Vilar2003}. The model assumes
that there is one main and one auxiliary operator, and transcription
occurs if and only if the main operator is free. Given these assumptions,
they showed that the repression is given by the expression\begin{equation}
\mathcal{R}=1+\frac{Ne^{-\triangle G_{m}}+Ne^{-\triangle G_{m}-\triangle G_{a}-\triangle G_{l}}+N(N-1)e^{-\triangle G_{m}-\triangle G_{a}}}{1+Ne^{-\triangle G_{a}}},\label{eq:Vilar}\end{equation}
where $N$ is the number of repressor molecules per cell; $\triangle G_{m},\triangle G_{a}$
are the free energy changes (normalized by $RT$) due to binding of
the repressor to the main and auxiliary operator, respectively; and
$\triangle G_{l}$ is the free energy change of loop formation. Equation~(\ref{eq:Vilar})
captures the repression of pairs of operators for suitable values
of $N$, $\triangle G_{m},$ $\triangle G_{a}$ and $\triangle G_{l}$.
Furthermore, the term, $Ne^{-\triangle G_{m}-\triangle G_{a}-\triangle G_{l}}$,
explains why DNA loops are so stable despite the weak repressor-operator
binding. If the magnitude of the looping free energy, $\left|\triangle G_{l}\right|$,
is sufficiently large, it can overcome the effect of small $\left|\triangle G_{m}\right|,\left|\triangle G_{a}\right|$.

\begin{figure}
\noindent \begin{centering}\subfigure[]{\includegraphics[width=2.5in,keepaspectratio]{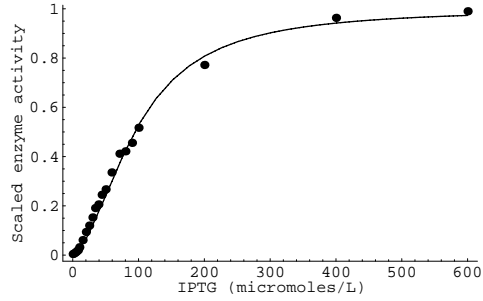}}\par\end{centering}

\noindent \begin{centering}\subfigure[]{\includegraphics[width=2.5in,keepaspectratio]{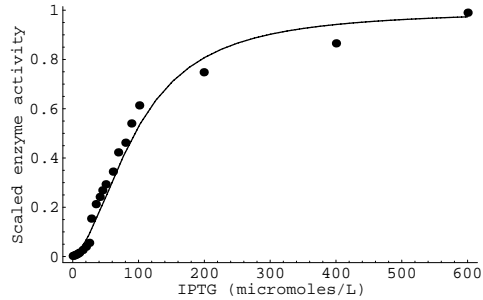}}\par\end{centering}

\noindent \begin{centering}\subfigure[]{\includegraphics[width=2.5in,keepaspectratio]{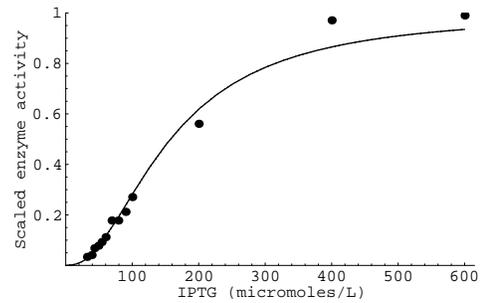}}\par\end{centering}

\caption{\label{f:OehlerInductionData}DNA looping increases the cooperativity
of the induction curve~\citep{Oehler2006}. (a,b) The induction curves
for cells containing (a) no auxiliary operators, and (b)~mutant dimeric
repressor. The data was fitted with eq.~(\ref{eq:Case1T}) and the
parameter values estimated by Oehler et al. (c)~The induction curve
for cells containing all three operators and tetrameric repressor.
The data was fitted with eq.~(\ref{eq:Case2T}) and the parameter
values in Table~\ref{t:Param}.}
\end{figure}

The above discussion shows that DNA looping strongly influences the
magnitude of the repression (observed in the absence of the inducer).
However, insofar as the formulation of dynamic models is concerned,
it is of more interest to ask if DNA looping influences the kinetics
of induction (observed in the presence of the inducer). It turns out
that this is indeed the case. Recently, Oehler et al compared the
induction kinetics in the absence and presence of DNA looping~\citep{Oehler2006}.
They abolished DNA looping by deleting the DNA encoding the auxiliary
operators, or mutating the DNA encoding the oligomerization domain
of the repressor (this results in the production of mutant dimers
that cannot form the tetrameric structure necessary for DNA looping).
In both cases, the induction kinetics were hyperbolic at all but the
smallest inducer concentrations (Figs.~\ref{f:OehlerInductionData}a,b).
In sharp contrast, the kinetics were strongly sigmoidal in the presence
of DNA looping (Fig.~\ref{f:OehlerInductionData}c). The authors
concluded that the {}``sigmoidality of the induction curve of the
wt lac system reflects cooperative repression through DNA loop formation.''

These experiments show that DNA looping massively amplifies the cooperativity
of the induction kinetics. The goal of this work is to understand
this phenomenon quantitatively. It is clear that we cannot appeal
to the Yagil \& Yagil model, since it does not account for the auxiliary
operators and the attendant DNA looping. Here, we formulate a model
of \emph{lac} induction taking due account of both features. We find
that

\begin{enumerate}
\item In the absence of DNA looping, the kinetics are formally similar to
eq.~(\ref{eq:YagilGeneral}), the general form the Yagil \& Yagil
model. However, in the presence of DNA looping, the kinetics are significantly
more cooperative.
\item In wild-type cells, they depend on powers of $x$ as high as $x^{4}$.
The cooperativity increases markedly because looped repressor-operator
complexes are very sensitive to the inducer concentrations.
\item If the repressor is overexpressed in wild-type cells, the kinetics
become even more cooperative --- they depend on powers of $x$ up
to $x^{6}$. Under these conditions, multiple repressors are bound
to the operons. These multi-repressor operons are even more sensitive
to inducer concentrations than operons with one repressor typically
found in wild-type cells.
\item The model provides good fits to the induction curves for 6 different
strains of \emph{E. coli}. More importantly, however, the model implies
the existence of specific scaling relations between looped and non-looped
complexes. These relations, which can be tested by gel electrophoresis,
provide a more stringent test of the model.
\end{enumerate}

\section{The model}

We begin by enumerating all possible states of the \emph{lac} operon.
We then define the transcription rate in terms of the concentrations
of the particular states that allow transcription. Finally, we derive
the governing equations that determine the concentrations of these
states as a function of the inducer concentration.

\begin{figure}
\begin{centering}\includegraphics[width=4.5in,keepaspectratio]{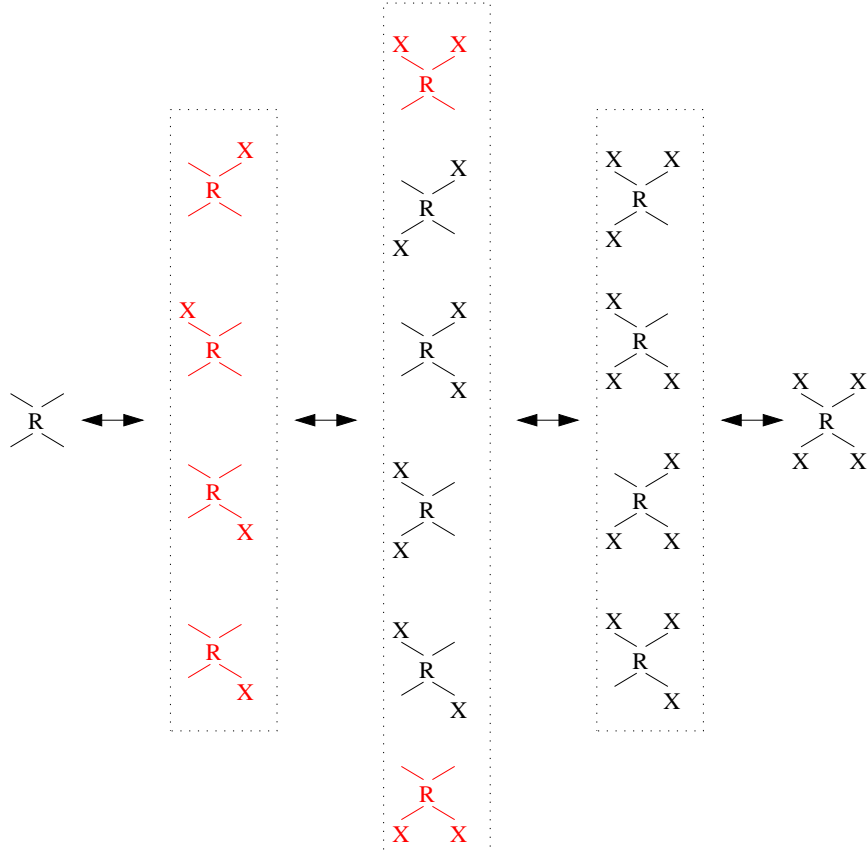}\par\end{centering}

\caption{\label{f:RepressorStates}All possible states of repressor-inducer
complexes. Here, $R$ and $X$ denote the repressor tetramer and inducer,
respectively. The free repressor is on the left. Repressor-inducer
complexes containing one free dimer are shown in red.}
\end{figure}

\subsection{States of the \emph{lac} operon}

We denote the \emph{free} repressor (i.e., repressor not bound to
an inducer or operator) and its concentration by $R$ and $r$, respectively.
Since the free repressor has 4 inducer binding sites, there are 15
possible repressor-inducer complexes (Fig.~\ref{f:RepressorStates}).
We denote the concentrations of repressor-inducer complexes containing
1, 2, 3, and 4 inducer molecules by $r_{1}$, $r_{2}$, $r_{3}$,
and $r_{4}$, respectively.

We assume that a repressor dimer can bind to an operator if and only
if it contains no inducer. It follows that:

\begin{enumerate}
\item In addition to the free repressor, there are six repressor-inducer
complexes that can bind to the operator (shown in red in Fig.~\ref{f:RepressorStates}).
We denote any inducer-bound repressor with one free dimer by $R'$,
and the total concentration of such complexes by $r'$.
\item Although both $R$ and $R'$ can bind to an operator, only operator-bound
$R$ can form DNA loops (Fig.~\ref{f:DNAloopFormation}). Operator-bound
$R'$ lacks the free dimer necessary for forming a DNA loop (Fig.~\ref{fig:StructureOfRepressor}).
\end{enumerate}
These two facts will be crucial for explaining the influence of DNA
looping on the induction kinetics.

\begin{figure}
\begin{centering}\includegraphics[width=4.5in,keepaspectratio]{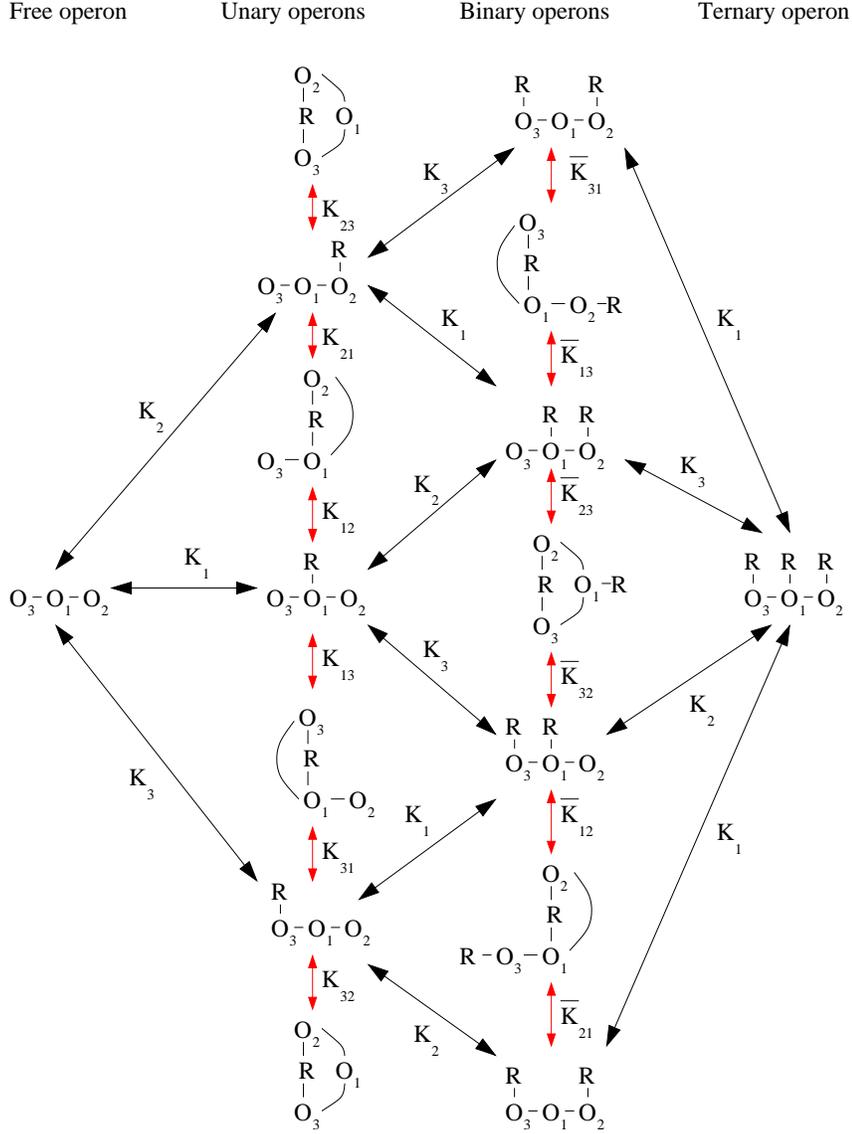}\par\end{centering}

\caption{\label{f:OperatorStates}All possible states of the \emph{lac} operon
when only free repressor is permitted to bind to the operators. The
black arrows show the reactions which a repressor binds to operator
$O_{i}$, and $K_{i}$ denotes the corresponding association constant.
The red arrows show the reactions in which a repressor-bound operator,
$O_{i}-R$, forms a loop by binding to a free operator $O_{j}$; $K_{ij}$
and $\bar{K}_{ij}$ denote the corresponding association constants
for unary and binary operons, respectively.}
\end{figure}

The \emph{lac} operon can be in numerous states. There are 14 possible
states if we assume that only $R$ can bind to an operator (Fig.~\ref{f:OperatorStates}).
Several additional states are feasible because $R'$ can also bind
to an operator. To enumerate these states systematically, it is convenient
to classify them based on the number of repressors bound to an operon.
We shall refer to operons containing 0, 1, 2, and 3 repressors as
\emph{free}, \emph{unary}, \emph{binary}, and \emph{ternary} operons,
respectively.

The operon can be free in only 1 way. We denote the concentration
of free operons by $o$.

Unary operons can exist in 9 different states. Six of these correspond
to states in which either $R$ or $R'$ is bound to one of the operators,
say, $O_{i}$. We denote the concentrations of these states by $o_{i}$
and $o_{i'}$, respectively. The remaining three states are obtained
because free repressor bound to $O_{i}$ can interact with another
operator, $O_{j}$, to form a DNA loop. We denote the concentration
of such looped states by $o_{\widehat{ij}}$. For example, $o_{\widehat{12}}$
denotes the concentration of the looped state obtained when a free
repressor bound to $O_{1}$ interacts with $O_{2}$, or a free repressor
bound to $O_{2}$ interacts with $O_{1}$. These definitions imply
that \begin{equation}
u=o_{1}+o_{2}+o_{3}+o_{1'}+o_{2'}+o_{3'}+o_{\widehat{31}}+o_{\widehat{12}}+o_{\widehat{32}},\label{eq:u}\end{equation}
where $u$ denotes the total concentration of the unary operons.

Binary operons can exist in 18 different states. Twelve of these correspond
to the states obtained when $R$ or $R'$ bind to any two of the 3
operators. We denote the concentrations of such states by $o_{ij}$,
$o_{i'j}$, $o_{ij'}$, $o_{i'j'}$, where the indices $i,j$ represent
the two operators to which $R$ or $R'$ are bound, and the symbol
$'$ above an index indicates that $R'$, rather than $R$, is bound
to the corresponding operator. The remaining 6 complexes are the looped
states obtained when the free dimer of an operator-bound free repressor
interacts with another free operator. We denote the concentration
of such looped complexes by overlaying the symbol $\widehat{}$ on
the subscripts representing the two interacting operators. For example,
$o_{3\widehat{12}}$ and $o_{3'\widehat{12}}$ denote the concentrations
of the states in which operator 3 is bound to $R$ and $R'$ respectively,
and operators 1,~2 interact by looping. It follows that \begin{align}
b & =\left(o_{31}+o_{3'1}+o_{31'}+o_{3'1'}\right)+\left(o_{12}+o_{1'2}+o_{12'}+o_{1'2'}\right)\nonumber \\
 & \quad+\left(o_{32}+o_{3'2}+o_{32'}+o_{3'2'}\right)\nonumber \\
 & \quad+\left(o_{\widehat{31}2}+o_{\widehat{31}2'}+o_{3\widehat{12}}+o_{3'\widehat{12}}+o_{\widehat{312}}+o_{\widehat{31'2}}\right),\label{eq:b}\end{align}
where $b$ denotes the total concentration of the binary operons.

Ternary operons can exist in 9 possible states, none of which are
looped because loops cannot form in ternary operons. The concentrations
of these states are denoted by $o_{\cdot\cdot\cdot}$, where each
$\cdot$ contains an integer of the form $i$ or $i'$ indicating
whether $R$ or $R'$ is bound to the $i$-th operator. Evidently\begin{equation}
t=\left(o_{312}+o_{31'2}+o_{312'}+o_{31'2'}\right)+\left(o_{3'12}+o_{3'1'2}+o_{3'12'}+o_{3'1'2'}\right),\label{eq:t}\end{equation}
where $t$ denotes the total concentration of the ternary complexes.

\subsection{Transcription rate}

Oehler et al have postulated that:

\begin{enumerate}
\item Binding of the repressor to $O_{1}$ blocks transcription by occluding
RNA polymerase~\citep[Chap.~1.18]{Muller-Hill}.
\item Binding of the repressor to $O_{2}$ has no effect on the transcription
rate. This is not because the repressor rarely binds to $O_{2}$:
Even if the repressor is overexpressed 90-fold, $O_{2}$-containing
cells show no measurable repression~\citep[Table~I]{Oehler1990}.
This suggests that $O_{2}$-bound repressor cannot obstruct the movement
of RNA polymerase.
\item Binding of the repressor to $O_{3}$ does not block transcription.
It merely reduces (deactivates) the transcription rate by preventing
CAP from binding to the repressor.\\
This hypothesis is based on the following argument. If repressor-bound
$O_{3}$ blocked transcription, the repression in $O_{3}$-containing
cells would increase monotonically with the repressor level. However,
if the repressor is overexpressed in these cells, the repression saturates
at 25~\citep[p.~3351]{Oehler1994}.
\end{enumerate}
These postulates imply that the transcription rate is proportional
to \[
T\equiv\frac{o}{o_{t}}+\left(\frac{o_{2}}{o_{t}}+\frac{o_{2'}}{o_{t}}\right)+d\left(\frac{o_{3}}{o_{t}}+\frac{o_{3'}}{o_{t}}+\frac{o_{32}}{o_{t}}+\frac{o_{3'2}}{o_{t}}+\frac{o_{32'}}{o_{t}}+\frac{o_{3'2'}}{o_{t}}+\frac{o_{\widehat{32}}}{o_{t}}\right),\]
where $d<1$ is a parameter accounting for deactivation of transcription
by repressor-bound $O_{3}$.

\subsection{Governing equations}

To determine the concentrations of the various states, we assume that

\begin{enumerate}
\item The total concentrations of the repressor and operator, denoted $r_{t}$
and $o_{t}$, are constant.
\item The system is in thermodynamic equilibrium, and satisfies the principle
of detailed balance (i.e., the net rate of every reaction is zero).
\item The binding of $R$ or $R'$ to an operator does not affect the affinity
of the remaining free operators for $R$ and $R'$. Hence, one can
define $K_{i}$ and $K_{i'}$ as the association constants for the
binding of $R$ and $R'$ to $O_{i}$, regardless of the state of
the remaining operators. Evidently, $K_{i'}=K_{i}/2$, since $R$
contains two inducer-free dimers, both of which can bind to $O_{i}$,
whereas $R'$ contains only 1 inducer-free dimer.\\
We denote the association constants for formation of unary and binary
loops by $K_{ij}$ and $\bar{K}_{ij}$, respectively (Fig.~\ref{f:OperatorStates}).
\item All four inducer-binding sites on the repressor are identical and
independent. We denote the association constant for binding of an
inducer to any one of these sites by $K_{x}$.
\end{enumerate}
Assumption 1 implies the \emph{conservation} relations\begin{align}
\left(r+r_{1}+r_{2}+r_{3}+r_{4}\right)+u+2b+3t & =r_{t},\label{eq:rConsvn}\\
o+u+b+t & =o_{t},\label{eq:oConsvn}\end{align}
where the factors 2 and 3 in (\ref{eq:rConsvn}) account for the fact
that the binary and ternary operons contain 2 and 3 repressors, respectively.

Assumptions 2 and 3 yield the \emph{equilibrium} relations

\begin{center}$\begin{array}{ccc}
o_{i}=K_{i}or & o_{i'}=\frac{1}{2}K_{i}or' & o_{\widehat{ij}}=K_{ij}K_{i}or=K_{ji}K_{j}or,\\
o_{ij}=K_{i}K_{j}or^{2} & o_{i'j}=o_{ij'}=\frac{1}{2}K_{i}K_{j}orr' & o_{i'j'}=\frac{1}{4}K_{i}K_{j}o\left(r'\right)^{2},\end{array}$\par\end{center}

and

\begin{center}$\begin{array}{cc}
o_{\widehat{31}2}=K_{2}K_{3}\bar{K}_{31}or^{2}=K_{2}K_{1}\bar{K}_{13}or^{2} & o_{\widehat{31}2'}=\frac{1}{2}K_{2}K_{3}\bar{K}_{31}orr'=\frac{1}{2}K_{2}K_{1}\bar{K}_{13}orr',\\
o_{3\widehat{12}}=K_{3}K_{1}\bar{K}_{12}or^{2}=K_{3}K_{2}\bar{K}_{21}or^{2} & o_{3'\widehat{12}}=\frac{1}{2}K_{3}K_{1}\bar{K}_{12}orr'=\frac{1}{2}K_{3}K_{2}\bar{K}_{21}orr',\\
o_{\widehat{312}}=K_{1}K_{3}\bar{K}_{32}or^{2}=K_{1}K_{2}\bar{K}_{23}or^{2} & o_{\widehat{31'2}}=\frac{1}{2}K_{1}K_{3}\bar{K}_{32}or^{2}=\frac{1}{2}K_{1}K_{2}\bar{K}_{23}or^{2},\\
o_{312}=K_{1}K_{2}K_{3}or^{3} & o_{3'12}=o_{31'2}=o_{312'}=\frac{1}{2}K_{1}K_{2}K_{3}or^{2}r',\\
o_{3'1'2'}=\frac{1}{8}K_{1}K_{2}K_{3}o\left(r'\right)^{3} & o_{31'2'}=o_{3'12}=o_{3'1'2}=\frac{1}{4}K_{1}K_{2}K_{3}or\left(r'\right)^{2},\end{array}$\par\end{center}

where the concentrations of the looped species have two representations
(e.g., $o_{\widehat{ij}}=K_{ij}K_{i}or=K_{ji}K_{j}or$) because these
species can be formed by two different pathways (Fig.~\ref{f:OperatorStates}).%
\footnote{Since the system is at equilibrium, thermodynamics demands that the
free energy changes of the two pathways be the same ($K_{i}K_{ij}=K_{j}K_{ji}$
in the above example). The principle of detailed balance ensures that
these thermodynamic constraints are satisfied~\citep{Feinberg1989}.%
}

These equilibrium relations imply that eqs.~(\ref{eq:u}--\ref{eq:t})
can be rewritten as\begin{align}
u & =o\left[\left(K_{1}+K_{2}+K_{3}\right)\left(r+\frac{r'}{2}\right)+\left(K_{1}K_{12}+K_{1}K_{13}+K_{2}K_{23}\right)r\right],\label{eq:uEq}\\
b & =o\left[\left(K_{1}K_{2}+K_{1}K_{3}+K_{2}K_{3}\right)\left(r+\frac{r'}{2}\right)^{2}\right.\label{eq:bEq}\\
 & \qquad\qquad+K_{1}\left(K_{2}\bar{K}_{23}+K_{2}\bar{K}_{13}+K_{3}\bar{K}_{12}\right)\left(r+\frac{r'}{2}\right),\nonumber \\
t & =o\cdot K_{1}K_{2}K_{3}\left(r+\frac{r'}{2}\right)^{3}.\label{eq:tEq}\end{align}
Assumptions 2 and 4 imply that the total concentration of all the
complexes shown in Fig.~\ref{f:RepressorStates} is given by \begin{align}
r+r_{1}+r_{2}+r_{3}+r_{4} & =r\left(1+K_{x}x\right)^{4},\label{eq:AllComplexes}\end{align}
and the total concentration of repressor-inducer complexes with one
free dimer is \begin{align}
r' & =r\left(4K_{x}x+2K_{x}^{2}x^{2}\right)\Rightarrow r+\frac{r'}{2}=r\left(1+K_{x}x\right)^{2}.\label{eq:FreeDimerComplexes}\end{align}
These two equations follow immediately from statistical thermodynamic
theory~\citep{Ackers1982}.

Substituting (\ref{eq:uEq}--\ref{eq:FreeDimerComplexes}) in (\ref{eq:rConsvn})--(\ref{eq:oConsvn})
yields the two governing equations\begin{align}
r\left(1+K_{x}x\right)^{4}\nonumber \\
+or\left[\left(K_{1}+K_{2}+K_{3}\right)\left(1+K_{x}x\right)^{2}+K_{1}K_{12}+K_{1}K_{13}+K_{2}K_{23}\right]\nonumber \\
+2or^{2}\left[\left(K_{1}K_{2}+K_{1}K_{3}+K_{2}K_{3}\right)\left(1+K_{x}x\right)^{4}\right.\nonumber \\
\left.+K_{1}\left(\bar{K}_{23}+\bar{K}_{13}+\bar{K}_{12}\right)\left(1+K_{x}x\right)^{2}\right]+3or^{3}\left(1+K_{x}x\right)^{6}K_{1}K_{2}K_{3} & =r_{t},\label{eq:r}\\
o+or\left[\left(K_{1}+K_{2}+K_{3}\right)\left(1+K_{x}x\right)^{2}+K_{1}K_{12}+K_{1}K_{13}+K_{2}K_{23}\right]\nonumber \\
+or^{2}\left[\left(K_{1}K_{2}+K_{1}K_{3}+K_{2}K_{3}\right)\left(1+K_{x}x\right)^{4}\right.\nonumber \\
\left.+K_{1}\left(\bar{K}_{23}+\bar{K}_{13}+\bar{K}_{12}\right)\left(1+K_{x}x\right)^{2}\right]+3or^{3}K_{1}K_{2}K_{3}\left(1+K_{x}x\right)^{6} & =o_{t},\label{eq:o}\end{align}
containing the 3 variables, $r,o,x$.

The equilibrium relations imply that\begin{align*}
T & =\frac{o}{o_{t}}\left[1+K_{2}r\left(1+K_{x}x\right)^{2}+d\left\{ K_{3}r\left(1+K_{x}x\right)^{2}\right.\right.\\
 & \qquad\qquad\left.\left.+K_{2}K_{3}r^{2}\left(1+K_{x}x\right)^{4}+K_{2}K_{23}r\right\} \right].\end{align*}
Eqs.~(\ref{eq:r})--(\ref{eq:o}) yield $o$ and $r$ as a function
of $x$, which can be substituted in the above expression to obtain
$T$ as a function of the inducer concentration.

\subsection{Scaled equations}

It is convenient to define the dimensionless variables\[
\rho\equiv\frac{r}{r_{t}},\nu\equiv\frac{o}{o_{t}},\chi\equiv K_{x}x,\]
and the dimensionless parameters\begin{align*}
\kappa_{i} & \equiv K_{i}r_{t},\; i=1,2,3,\\
\alpha_{1} & \equiv\kappa_{1}+\kappa_{2}+\kappa_{3},\\
\widehat{\alpha}_{1} & \equiv\kappa_{1}K_{12}+\kappa_{1}K_{13}+\kappa_{2}K_{23},\\
\alpha_{2} & \equiv\kappa_{1}\kappa_{2}+\kappa_{1}\kappa_{3}+\kappa_{2}\kappa_{3},\\
\widehat{\alpha}_{2} & \equiv\kappa_{1}\left(\kappa_{2}\bar{K}_{23}+\kappa_{2}\bar{K}_{13}+\kappa_{3}\bar{K}_{12}\right),\\
\alpha_{3} & \equiv\kappa_{1}\kappa_{2}\kappa_{3},\\
\omega & \equiv\frac{o_{t}}{r_{t}}.\end{align*}
The transcription rate is then proportional to\begin{align}
T & =\nu\left[1+\kappa_{2}\rho(1+\chi)^{2}+d\left\{ \kappa_{3}\rho(1+\chi)^{2}\right\} \right.\nonumber \\
 & \qquad\qquad\qquad\left.\left.+\kappa_{2}\kappa_{3}\rho^{2}(1+\chi)^{4}+\kappa_{2}K_{23}\rho\right\} \right]\label{eq:tRate}\end{align}
and eqs.~(\ref{eq:r})--(\ref{eq:o}) become\begin{alignat}{1}
\rho\left(1+\chi\right)^{4}+\omega\nu\left[\rho f_{1}(\chi)+2\rho^{2}f_{2}(\chi)+3\rho^{3}f_{3}(\chi)\right] & =1,\label{eq:rho}\\
\nu\left[1+\rho f_{1}(\chi)+\rho^{2}f_{2}(\chi)+\rho^{3}f_{3}(\chi)\right] & =1,\label{eq:nu}\end{alignat}
where\begin{align*}
f_{1}(\chi) & \equiv\alpha_{1}\left(1+\chi\right)^{2}+\widehat{\alpha}_{1},\\
f_{2}(\chi) & \equiv\alpha_{2}\left(1+\chi\right)^{4}+\widehat{\alpha}_{2}\left(1+\chi\right)^{2},\\
f_{3}(\chi) & \equiv\alpha_{3}\left(1+\chi\right)^{6}.\end{align*}
As we show below, the parameters, $\alpha_{i}$ and $\widehat{\alpha}_{i}$,
are related to the repression due to repressor-operator binding and
DNA looping, respectively. The parameter, $\omega$, is typically
quite small. In wild-type \emph{Escherichia coli}, $\omega\approx0.2$
since each cell contains 10 repressor molecules and no more than 2
operators~\citep[Chap.~3.2]{Muller-Hill}. In many experiments, the
repressor is overexpressed (>50 molecules per cell), so that $\omega<0.02$.

\section{Results\label{s:Results}}

In what follows, we shall determine the values of $\alpha_{i}$ and
$\widehat{\alpha}_{i}$ by appealing to the repression data. It is
therefore useful to express the repression in terms of the model.

To this end, we begin by observing that during exponential growth
in the presence of IPTG and glycerol, the mass balance for $\beta$-galactosidase
yields\[
\frac{de}{dt}=r_{e}(x)-\left(r_{g}+k_{e}^{-}\right)e=0\Rightarrow e=\frac{r_{e}(x)}{r_{g}+k_{e}^{-}},\]
where $x$ is the concentration of IPTG, $r_{e}(x)$ is the corresponding
specific rate of $\beta$-galactosidase synthesis, $r_{g}$ is the
maximum specific growth rate on glycerol, and $k_{e}^{-}$ is the
rate constant for $\beta$-galactosidase degradation. Since $r_{g}+k_{e}^{-}$
is a fixed parameter, $e$ is proportional to $r_{e}$, and (\ref{eq:Rdefn})
becomes \[
\mathcal{R}=\frac{\left.r_{e}(x)\right|_{x\rightarrow\infty}}{r_{e}(0)}.\]
It follows from (\ref{eq:tRate}) that \begin{equation}
\mathcal{R}=\frac{1}{T(0)}=\frac{1}{\nu(0)\left[1+\kappa_{2}+d\left(\kappa_{3}+\kappa_{2}\kappa_{3}+\kappa_{2}K_{23}\right)\right]},\label{eq:R}\end{equation}
where we have assumed that $\rho(0)=1$, and at large inducer concentrations,
$\rho=0,\nu=1$.

Oehler et al measured the repression in the presence of various combinations
of operators (Table~\ref{t:RepressionData}). We shall distinguish
these cases by using subscripts to denote the particular combination
of operators being considered. Specifically, $\mathcal{R}_{i}$ will
denote the repression in cells containing only the $i$-th operator,
$\mathcal{R}_{ij}$ will denote the repression in cells containing
the $i$-th and $j$-th operators, and $\mathcal{R}_{312}$ will denote
the repression in cells containing all 3~operators.

We begin by considering the special cases in which there is no DNA
looping, and then proceed to the more general case that accounts for
DNA looping.

\subsection{No DNA looping}

In the experiments, DNA looping was abolished by deleting the auxiliary
operators or mutating the locus for the oligomerization domain of
the repressor. Here, we consider the first case. The case of mutant
dimers is discussed in Appendix~\ref{a:MutantDimers}.

In the absence of the auxiliary operators, $\kappa_{2}=\kappa_{3}=0$,
so that\begin{equation}
\alpha_{1}=\kappa_{1},\;\widehat{\alpha}_{1}=\alpha_{2}=\widehat{\alpha}_{1}=\alpha_{3}=0,\label{eq:Case1alpha1}\end{equation}
and eqs.~(\ref{eq:tRate})--(\ref{eq:nu}) become $T=\nu,$ and \begin{align}
\rho\left(1+\chi\right)^{4}+\omega\nu\rho\kappa_{1}\left(1+\chi\right)^{2} & =1,\label{eq:Case1rho}\\
\nu\left[1+\rho\kappa_{1}\left(1+\chi\right)^{2}\right] & =1.\label{eq:Case1nu}\end{align}
We can get $\nu(\chi)$ from these equations by eliminating $\rho$
and solving the resulting quadratic. However, this solution is cumbersome
and offers little insight. Instead, since $\omega$ is small, we appeal
to perturbation theory (Appendix~\ref{a:MolecularParameters}), which
formalizes the following physical argument.

Since the number of operons per cell is small compared to the number
of repressors per cell, one can assume, as a first approximation,
that the fraction of operon-bound repressors is negligibly small compared
to the fraction of free repressors, i.e., $\omega=0$. Equations~(\ref{eq:Case1rho})--(\ref{eq:Case1nu})
then yield the approximate zeroth-order solution\begin{align}
\rho_{0} & =\frac{1}{\left(1+\chi\right)^{4}},\label{eq:Case1rho0}\\
\nu_{0} & =\frac{1}{1+\kappa_{1}\rho_{0}\left(1+\chi\right)^{2}}.\label{eq:Case1nu0}\end{align}
To estimate the error of the approximation, we acknowledge that the
fraction of operon-bound repressors is small but not zero. We assume
furthermore that this fraction can be estimated by the expression,
$\omega\nu_{0}\rho_{0}\kappa_{1}\left(1+\chi\right)^{2}$, and solve
the resulting equations\begin{align*}
\rho\left(1+\chi\right)^{4}+\omega\nu_{0}\rho_{0}\kappa_{1}\left(1+\chi\right)^{2} & =1,\\
\nu\left[1+\rho\kappa_{1}\left(1+\chi\right)^{2}\right] & =1,\end{align*}
to obtain the improved first-order \emph{}solution \begin{align}
\rho & =\rho_{0}\left[1-\omega\left(1-\nu_{0}\right)\right]+O(\omega^{2}),\label{eq:Case1rho1}\\
\nu & =\nu_{0}\left[1+\omega\left(1-\nu_{0}\right)^{2}\right]+O(\omega^{2}).\label{eq:Case1nu1}\end{align}
It follows from (\ref{eq:Case1nu1}) that the relative error of $\nu_{0}$
is approximately\[
\frac{\nu-\nu_{0}}{\nu}=\frac{\omega\left(1-\nu_{0}\right)^{2}}{1+\omega\left(1-\nu_{0}\right)^{2}}<\frac{\omega}{1+\omega}.\]
Since, $\omega\lesssim0.2$, the zeroth-order solution is accurate
to within $100\omega/(1+\omega)\approx15$\% in wild-type cells, and
even more accurate in repressor-overexpressed cells. Henceforth, we
shall assume that eqs.~(\ref{eq:Case1rho0})--(\ref{eq:Case1nu0})
are a good approximation to the exact solution, so that \begin{equation}
T(\chi)=\nu(\chi)\approx\frac{1}{1+\kappa_{1}/\left(1+\chi\right)^{2}},\label{eq:Case1T}\end{equation}
which is formally identical to eq.~(\ref{eq:YagilGeneral}) with
$K_{x,1}=2K_{x},$ $K_{x,1}K_{x,2}=K_{x}^{2}$, the special case of
the Yagil \& Yagil model corresponding to identical and independent
inducer-binding sites~\citep[p~19]{yagil71}.

It follows from (\ref{eq:Case1T}) that induction is cooperative even
in the absence of DNA looping. Indeed, since $T(\chi)$ has a unique
inflection point at $\chi=\sqrt{\alpha_{1}/3}-1$, $T=1/4$, the kinetics
are cooperative for all inducer concentrations such that $0\le T\le1/4$.
In the particular case of Fig.~\ref{f:OehlerInductionData}a, the
kinetics are cooperative for all inducer concentrations in the range
0--50~$\mu$M, which is significantly higher than the 0--5~$\mu$M
range reported in~\citealp{Oehler2006}, based upon visual inspection
of the curve.

\begin{figure}
\noindent \begin{centering}\subfigure[]{\includegraphics[width=2.5in]{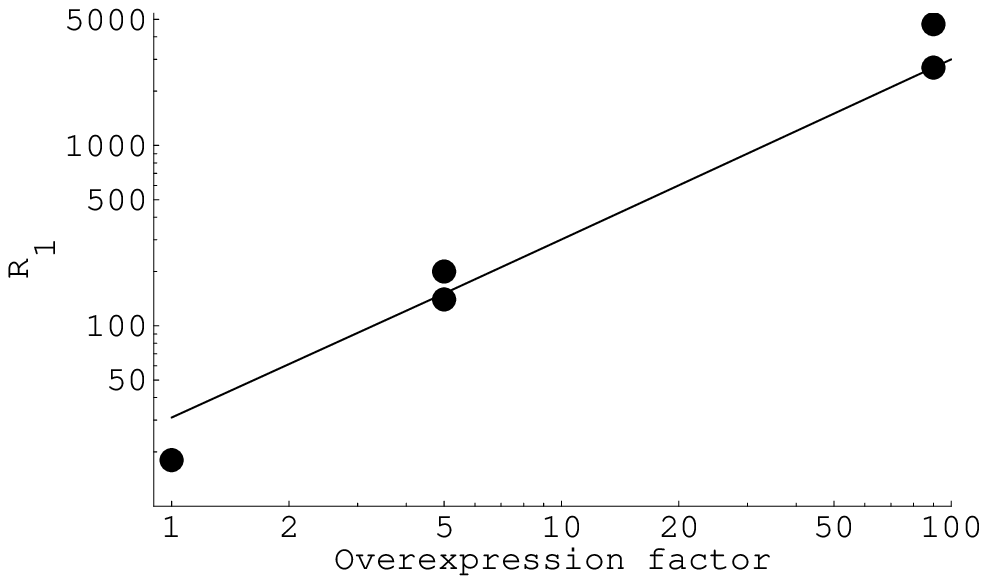}}\hspace*{0.3in}\subfigure[]{\includegraphics[width=2.5in]{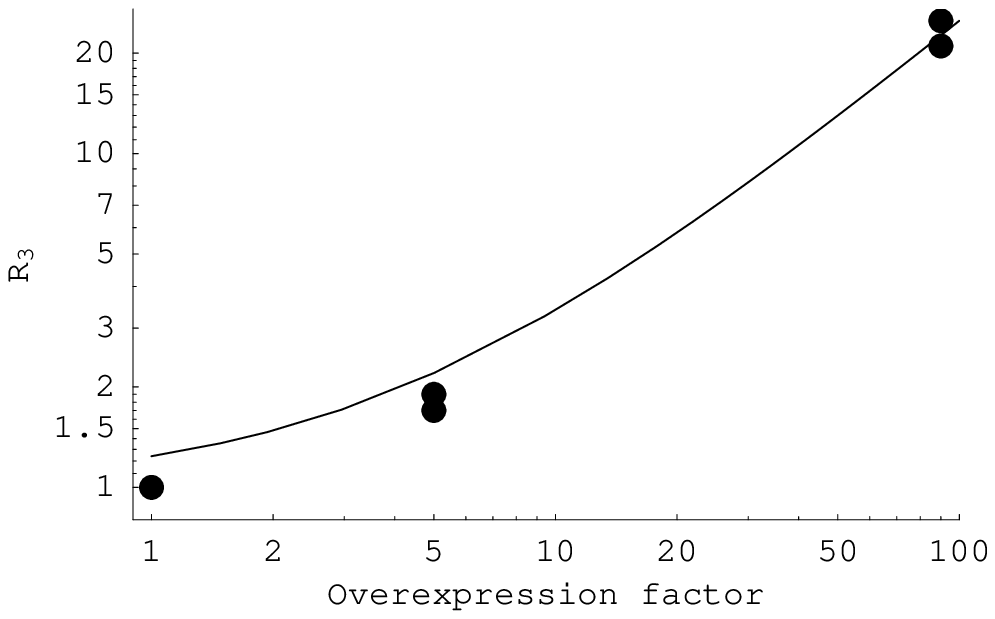}}\par\end{centering}

\caption{\label{f:OneOperatorRepression}Estimation of $\kappa_{1}$ and $\kappa_{3}$
by fitting the repression data from \citealp[Table I]{Oehler1990},
and \citealp[Fig.~1]{Oehler1994}, to eqs.~(\ref{eq:R1}--\ref{eq:R3}).}
\end{figure}

The parameters, $\kappa_{1}$, $K_{x1}$, can be estimated from the
induction curve by observing that (\ref{eq:Case1T}) implies \[
\sqrt{\frac{T}{1-T}}=\frac{1}{\sqrt{\kappa_{1}}}+\frac{K_{x}}{\sqrt{\kappa_{1}}}x.\]
If the model is correct, a plot of $\sqrt{T/(1-T)}$ vs $x$ will
be a straight line, and $\kappa_{1},K_{x}$ can be estimated from
the slope and $y$-intercept. The induction curve shown in Fig.~\ref{f:OehlerInductionData}a
yields a straight line with $\kappa_{1}=227$ and $K_{x}^{-1}=6.7$~$\mu\textnormal{M}$
\citep[Fig.~4B]{Oehler2006}.

The value of $\kappa_{1}$ can also be estimated from the repression
data. Indeed, it follows from (\ref{eq:Case1T}) that\begin{equation}
\mathcal{R}_{1}=\frac{1}{T(0)}=1+\kappa_{1}.\label{eq:R1}\end{equation}
Fitting the repression at various overexpression levels to this equation
yields $\kappa_{1}=30$ for wild-type cells (Fig.~\ref{f:OneOperatorRepression}a).
This is $\sim$7-fold lower than the value estimated above because
the induction curve was obtained with repressor-overexpressed cells.

Although eq.~(\ref{eq:Case1nu0}) was derived for cells containing
the main operator $O_{1}$, analogous expressions are obtained for
cells containing an auxiliary operator, i.e., $\nu=1/\left[1+\kappa_{i}/(1+\chi)^{2}\right]$
for $i=2,3$. Equation~(\ref{eq:R}) then implies that the repression
in $O_{3}$-containing cells is \[
\mathcal{R}_{3}=\frac{1+\kappa_{3}}{1+d\kappa_{3}},\]
which captures the deactivation effect noted by Oehler et al: $\mathcal{R}_{3}$
increases with the repressor level until it saturates $1/d$. However,
the data shows no evidence of this saturation even if the repressor
is overexpressed 90-fold (Fig.~\ref{f:OneOperatorRepression}b).
Nonlinear regression of the data using the above expression yields
the best-fit parameters, $d=0$ and $\kappa_{3}=0.24$ for wild-type
cells. It is conceivable that $d$ is positive, but so small that
$d\kappa_{3}\ll1$ for the overexpression levels shown in Fig.~\ref{f:OneOperatorRepression}b.
Henceforth, we shall assume that $d=0$, and \begin{equation}
\mathcal{R}_{3}=1+\kappa_{3},\label{eq:R3}\end{equation}
a relation that is valid up to an overexpression level of 90.%
\footnote{These estimates of $\kappa_{i}$ also provide good fits to the repression
data for dimers (Fig.~\ref{f:DimerRepression})%
}

Unlike $\kappa_{1}$ and $\kappa_{3}$, the parameter, $\kappa_{2}$,
cannot be calculated from the repression data for $O_{2}$-containing
cells because they show no repression even if the repressor is overexpressed
90-fold. This property is implicit in the model as well. Indeed, (\ref{eq:R})
implies that \[
\mathcal{R}_{2}=\frac{1}{\nu(0)(1+\kappa_{2})}=\frac{1+\kappa_{2}}{1+\kappa_{2}}=1,\]
 regardless of the repressor level. Evidently, this reflects the fact
that $O_{2}$-bound repressor does not block RNA polymerase.

\subsection{DNA looping}

In this case, the full system of eqs.~(\ref{eq:rho})--(\ref{eq:nu})
must be solved for $\nu$ and $\rho$. Perturbation theory yields
the zeroth-order solution\begin{align}
\rho_{0} & =\frac{1}{\left(1+\chi\right)^{4}},\label{eq:Case3rho0}\\
\nu_{0} & =\frac{1}{1+\rho_{0}f_{1}(\chi)+\rho_{0}^{2}f_{2}(\chi)+\rho_{0}^{3}f_{3}(\chi)}.\label{eq:Case3nu0}\end{align}
It is evident from (\ref{eq:Case3nu0}) that $\rho_{0}f_{1}(\chi)$,
$\rho_{0}^{2}f_{2}(\chi)$, and $\rho_{0}^{3}f_{3}(\chi)$ are the
concentrations of the unary, binary, and ternary operons, respectively,
relative to the concentration of the free operons. We shall constantly
appeal to this fact below.

The zeroth-order solution is a good approximation to the exact solution.
Indeed, the first order solution is given by (Appendix~\ref{a:MolecularParameters})\begin{align*}
\rho & =\rho_{0}\left(1-\omega\Omega_{0}\right)+O(\omega^{2}),\\
\nu & =\nu_{0}\left(1+\omega\Omega_{0}^{2}\right)+O(\omega^{2}),\end{align*}
where \begin{equation}
\Omega_{0}\equiv\frac{\rho_{0}f_{1}(\chi)+2\rho_{0}^{2}f_{2}(\chi)+3\rho_{0}^{3}f_{3}(\chi)}{1+\rho_{0}f_{1}(\chi)+\rho_{0}^{2}f_{2}(\chi)+3\rho_{0}^{3}f_{3}(\chi)},\label{eq:omega0}\end{equation}
and the relative error of $\nu_{0}$ is approximately\[
\frac{\nu-\nu_{0}}{\nu}=\frac{\omega\Omega_{0}^{2}}{1+\omega\Omega_{0}^{2}}.\]
The above interpretation of the terms, $\rho_{0}^{i}f_{i}(\chi),i=1,2,3$,
implies that $\Omega_{0}$ is the average number of repressors per
operon, and hence, can have any value between 0 and 3. At large inducer
concentrations, $\Omega_{0}\approx0$, and the error is guaranteed
to be vanishingly small. At low inducer concentrations, $\Omega_{0}$
can exceed 1, provided the fraction of binary and ternary operons
is sufficiently large. However, we show below that in wild-type cells,
$\Omega_{0}$ is close to 1 (Fig.~\ref{f:Overexpressed}a). In repressor-overexpressed
cells, $\Omega_{0}$ can approach 3, but $\omega$ is so small that
the relative error of $\nu_{0}$ does not exceed 20\% (Fig.~\ref{f:errorPlot}).
The zeroth-order solution is therefore a good approximation at all
repressor levels and inducer concentrations.

Substituting (\ref{eq:Case3rho0}) in (\ref{eq:Case3nu0}) and (\ref{eq:tRate})
with $d=0$ yields \begin{align}
\nu_{0} & =\frac{1}{1+\frac{\alpha_{1}}{\left(1+\chi\right)^{2}}+\frac{\widehat{\alpha}_{1}}{\left(1+\chi\right)^{4}}+\frac{\alpha_{2}}{\left(1+\chi\right)^{4}}+\frac{\widehat{\alpha}_{2}}{\left(1+\chi\right)^{6}}+\frac{\alpha_{3}}{\left(1+\chi\right)^{6}}},\label{eq:Case3nu0Final}\\
T & =\nu_{0}\left[1+\frac{\kappa_{2}}{(1+\chi)^{2}}\right],\label{eq:Case3T}\end{align}
which shows that in the presence of DNA looping, the induction rate
is formally different from (\ref{eq:Case1T}). It turns out, however,
that in wild-type \emph{lac}, the parameter values are such that several
terms in the above expresssions are negligibly small. To see this,
it is useful to define\begin{align}
\phi_{i} & (\chi)\equiv\frac{\alpha_{i}}{\left(1+\chi\right)^{2i}},\; i=1,2,3,\label{eq:phi}\\
\widehat{\phi}_{i} & (\chi)\equiv\frac{\widehat{\alpha}_{i}}{\left(1+\chi\right)^{2(i+1)}},\; i=1,2,\label{eq:phiHat}\end{align}
and rewrite (\ref{eq:Case3nu0Final}) as\begin{equation}
\nu_{0}=\frac{1}{1+\phi_{1}(\chi)+\widehat{\phi}_{1}(\chi)+\phi_{2}(\chi)+\widehat{\phi}_{2}(\chi)+\phi_{3}(\chi)}.\label{eq:nu0Phis}\end{equation}
Evidently, $\phi_{i}(\chi)$ and $\widehat{\phi}_{i}(\chi)$ are the
\emph{relative concentrations} of the non-looped and looped operons
containing $i$ repressors (measured relative to the concentration
of free operons). In particular, the parameters, $\alpha_{i}=\phi_{i}(0)$
and $\widehat{\alpha}_{i}=\widehat{\phi}_{i}(0)$, are the relative
concentrations of these operons in the absence of the inducer.

\begin{figure}
\noindent \begin{centering}\subfigure[]{\includegraphics[width=2.5in]{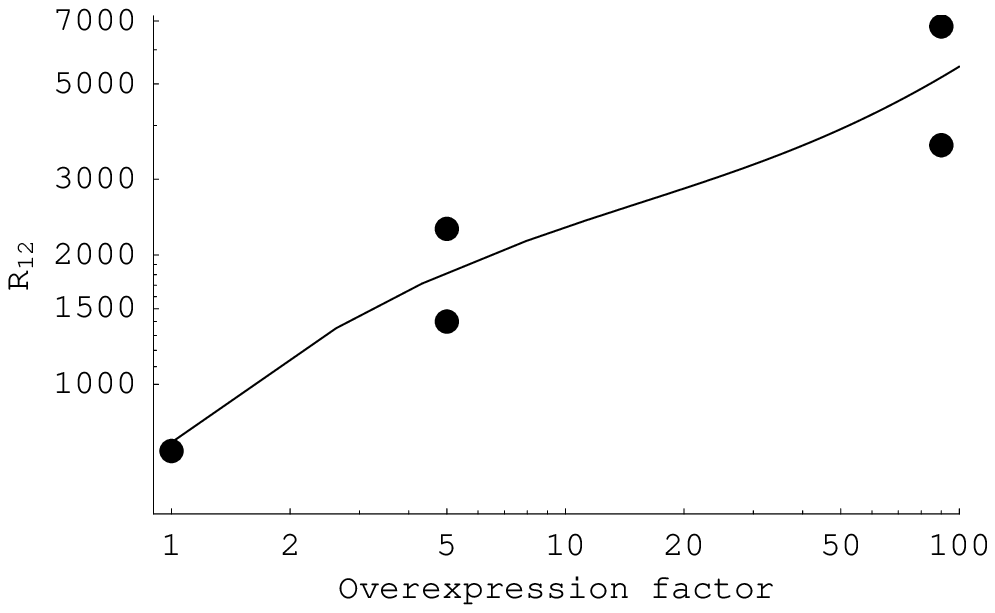}}\hspace*{0.3in}\subfigure[]{\includegraphics[width=2.5in]{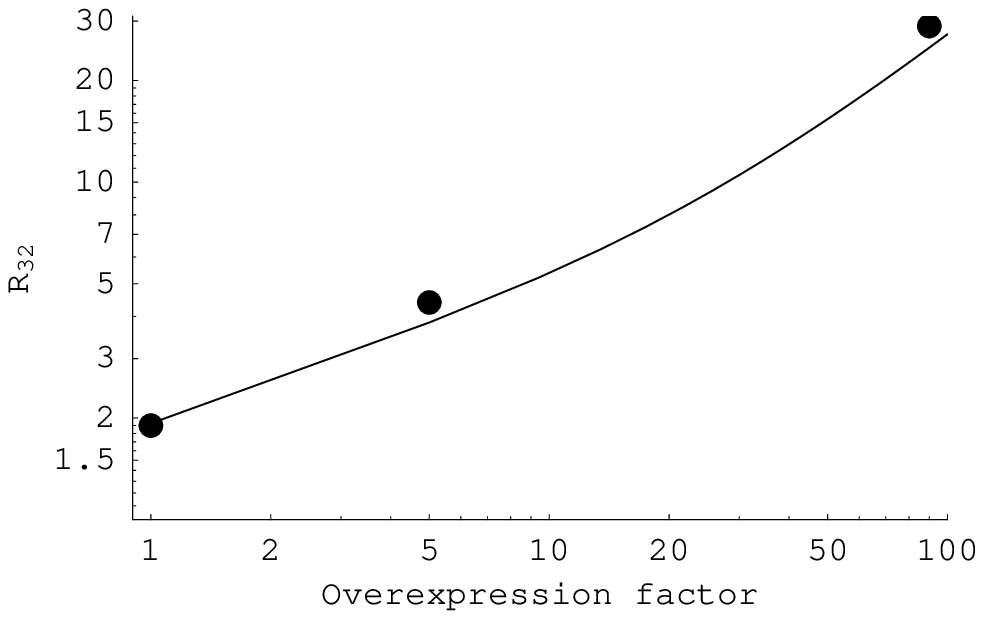}}\par\end{centering}

\noindent \begin{centering}\subfigure[]{\includegraphics[width=2.5in]{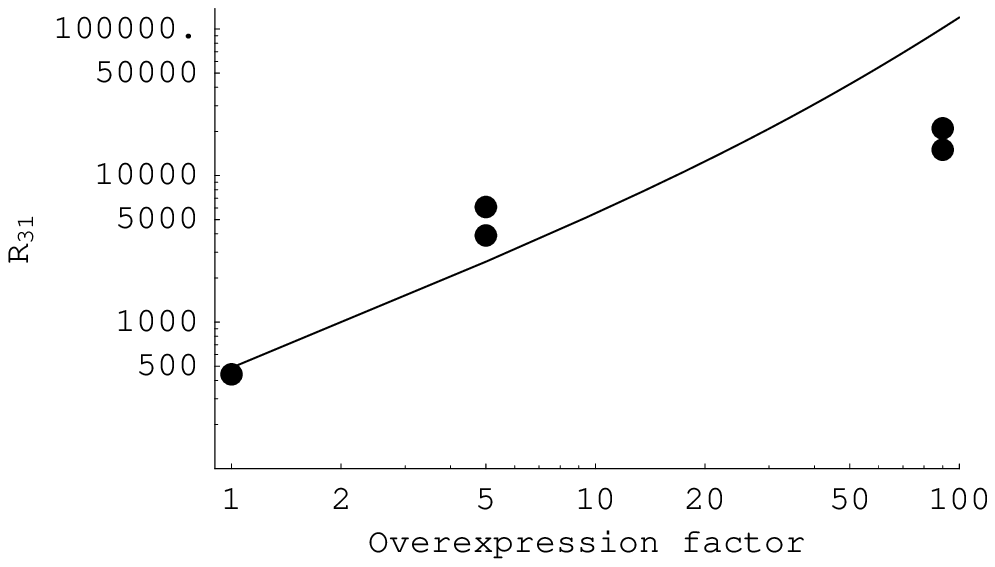}}\hspace*{0.3in}\subfigure[]{\includegraphics[width=2.5in]{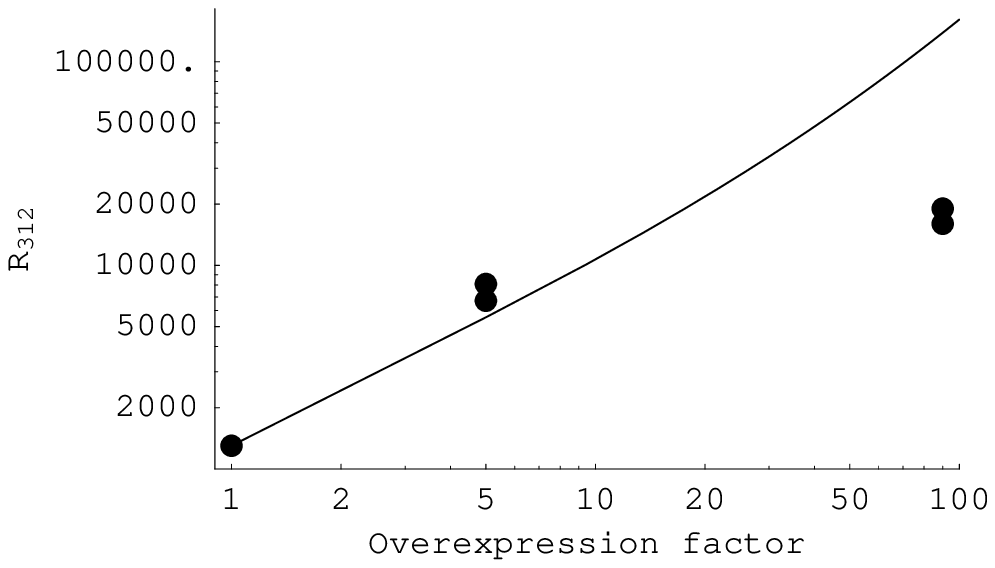}}\par\end{centering}

\caption{\label{f:MultiOperatorRepression}Estimation of $\kappa_{2}$, $K_{12}$,
$K_{13}$, and $K_{23}$ by fitting the repression data from \citealp[Table I]{Oehler1990}
and \citealp[Figs~4--5]{Oehler1994} to eqs.~(\ref{eq:R12}) and
(\ref{eq:R31}). The values of $\mathcal{R}_{31}$ and $\mathcal{R}_{312}$
at 90-fold overexpression are lower bounds for the repression. The
true repression levels are too high to be measured accurately.}
\end{figure}

We begin by determining the wild-type values of $\alpha_{i}$ and
$\widehat{\alpha}_{i}$. The above estimates of $\kappa_{1}$, $\kappa_{2}$,
and $\kappa_{3}$ imply that $\alpha_{1}=31$, $\alpha_{2}=19$, and
$\alpha_{3}=3$. To find the remaining parameters, $\widehat{\alpha}_{1},\widehat{\alpha}_{2}$,
observe that since\begin{equation}
\mathcal{R}_{312}=\frac{1}{T(0)}=\frac{1+\alpha_{1}+\widehat{\alpha}_{1}+\alpha_{2}+\widehat{\alpha}_{2}+\alpha_{3}}{1+\kappa_{2}},\label{eq:R312}\end{equation}
the repression in cells containing pairs of operators are given by
the expressions%
\footnote{Eq.~(\ref{eq:R12}) is the kinetic analog of eq.~(\ref{eq:Vilar})
derived from thermodynamic principles. %
} \begin{align}
\mathcal{R}_{12} & =\frac{1+\kappa_{1}+\kappa_{2}+\kappa_{1}K_{12}+\kappa_{1}\kappa_{2}}{1+\kappa_{2}},\label{eq:R12}\\
\mathcal{R}_{32} & =\frac{1+\kappa_{2}+\kappa_{3}+\kappa_{2}K_{23}+\kappa_{2}\kappa_{3}}{1+\kappa_{2}},\label{eq:R32}\\
\mathcal{R}_{31} & =1+\kappa_{1}+\kappa_{3}+\kappa_{1}K_{13}+\kappa_{1}\kappa_{3}.\label{eq:R31}\end{align}
Fitting the repression data obtained at various overexpression levels
to these equations yields the estimates, $\kappa_{2}=0.38$, $K_{12}=32$,
$K_{13}=15$, $K_{23}=2.5$ (Fig.~\ref{f:MultiOperatorRepression}),
which imply that\[
\widehat{\alpha}_{1}\equiv\kappa_{1}K_{12}+\kappa_{1}K_{13}+\kappa_{2}K_{23}=1420.\]
Since the measured value of $\mathcal{R}_{312}$ is 1300, eq.~(\ref{eq:R312})
implies that $\widehat{\alpha}_{2}=322$.

These parameter values imply that in wild-type cells, the induction
rate is much simpler than (\ref{eq:Case3T}). To see this, observe
that in the absence of the inducer, the relative concentrations of
binary and ternary operons are small compared to the relative concentrations
of free and unary operons, i.e., \begin{equation}
\alpha_{2}+\widehat{\alpha}_{2}+\alpha_{3}\ll1+\alpha_{1}+\widehat{\alpha}_{1}.\label{eq:UnaryDominance}\end{equation}
Now, eqs.~(\ref{eq:phi}--\ref{eq:phiHat}) imply that in the presence
of the inducer, the relative concentrations of the binary and ternary
operons decrease with the inducer concentration at a rate as fast,
or even faster, than the corresponding rate for the looped unary operons.
It follows that even in the presence of the inducer, the relative
concentrations of the binary and ternary operons remain negligibly
small compared to the relative concentrations of the unary and free
operons, i.e., the relation\[
\phi_{2}(\chi)+\widehat{\phi}_{2}(\chi)+\phi_{3}(\chi)\ll1+\phi(\chi)+\widehat{\phi}_{1}(\chi)\]
is true for all $\chi\ge0$. The fraction of free operons in wild-type
\emph{lac} is therefore well-approximated by the simpler expression\begin{align}
\nu_{0} & \approx\frac{1}{1+\phi_{1}(\chi)+\widehat{\phi}_{1}(\chi)}.\label{eq:Case2nu0}\end{align}
A similar argument shows that in the absence of the inducer, $\kappa_{2}/(1+\chi)^{2}$,
the relative concentration of $O_{2}$-bound operons, is 0.38, and
(\ref{eq:Case3T}) implies that almost 1/3 of the transcription occurs
from $O_{2}$-bound operons. However, $\kappa_{2}/(1+\chi)^{2}$ decreases
so rapidly with the inducer concentration that it is already below
0.2 at $\chi=0.5$. Thus, the transcription rate of wild-type \emph{lac}
is well-approximated by the expression\begin{equation}
T(\chi)\approx\nu_{0}(\chi)=\frac{1}{1+\alpha_{1}/\left(1+\chi\right)^{2}+\widehat{\alpha}_{1}/\left(1+\chi\right)^{4}}\label{eq:Case2T}\end{equation}
for all but a negligibly small range of inducer concentrations. This
expression is simpler than (\ref{eq:Case3T}), but formally different
from (\ref{eq:Case1T}). The physical reason for this will be discussed
shortly.

The parameter values also imply that in the absence of the inducer,
the relative concentrations of the free and non-looped unary operons
are negligibly small compared to relative concentration of looped
unary operons, i.e.,\[
1+\alpha_{1}\ll\widehat{\alpha}_{1}.\]
 It follows that in wild-type cells, the repression is exerted almost
entirely by the looped unary operons, i.e., \begin{equation}
\mathcal{R}_{312}\approx\frac{\widehat{\alpha}_{1}}{1+\kappa_{2}}\approx\frac{\kappa_{1}\left(K_{12}+K_{13}\right)}{1+\kappa_{2}}.\label{eq:Case2Repression}\end{equation}
This equation explains an important trend in Table~\ref{t:RepressionData}.
Specifically, the addition of only one of the auxiliary operators
to the main operator increases the repression dramatically (25- to
40-fold) because $K_{12},K_{13}\gg1$. However, addition of the second
auxiliary operator provokes no more than a 2- or 3-fold increase because
the magnitudes of $K_{12}$ and $K_{13}$ are comparable.

Comparison of (\ref{eq:Case1T}) and (\ref{eq:Case2T}) shows that
the induction kinetics are qualitatively different in the presence
of DNA looping precisely because $\widehat{\phi}_{1}(\chi)$ decreases
faster than $\phi_{1}(\chi)$. The physical reason for this is as
follows. Looped unary states can form only if free repressor binds
to an operator, whereas non-looped unary states can form if free or
inducer-bound repressor binds to an operator. More precisely, eqs.~(\ref{eq:uEq})
and (\ref{eq:FreeDimerComplexes}) show that the relative concentrations
of looped and non-looped unary operons are proportional to $r$ and
$r+r'/2=r(1+\chi)^{2}$, respectively. Since $r$ is proportional
to $(1+\chi)^{-4}$, $\widehat{\phi}_{1}(\chi)$ and $\phi_{1}(\chi)$
decrease at the rates $(1+\chi)^{-4}$ and $(1+\chi)^{-2}$, respectively.

\begin{figure}[t]
\noindent \begin{centering}\subfigure[]{\includegraphics[width=2.5in,keepaspectratio]{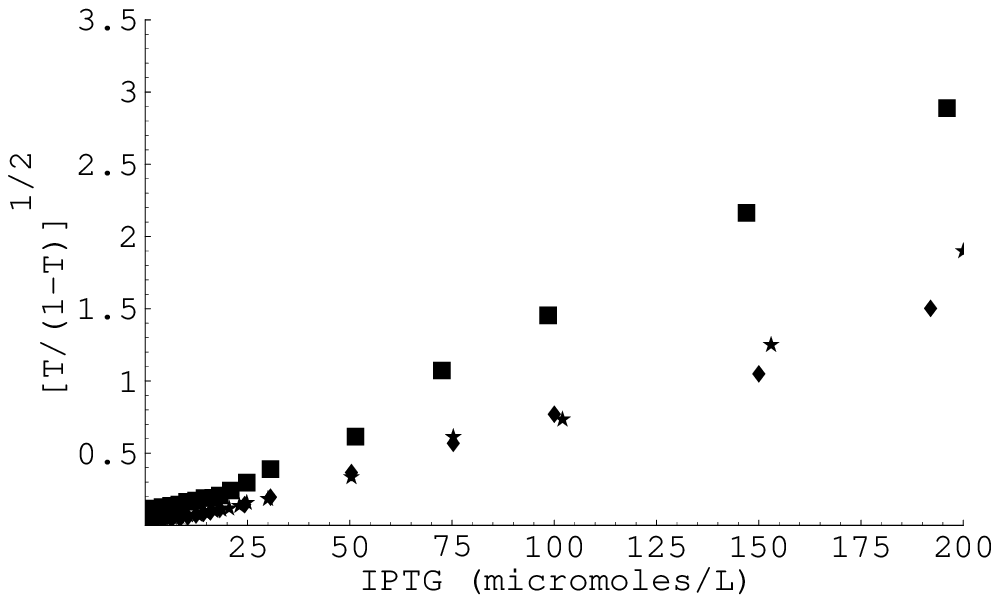}}\hspace*{0.3in}\subfigure[]{\includegraphics[width=2.5in,keepaspectratio]{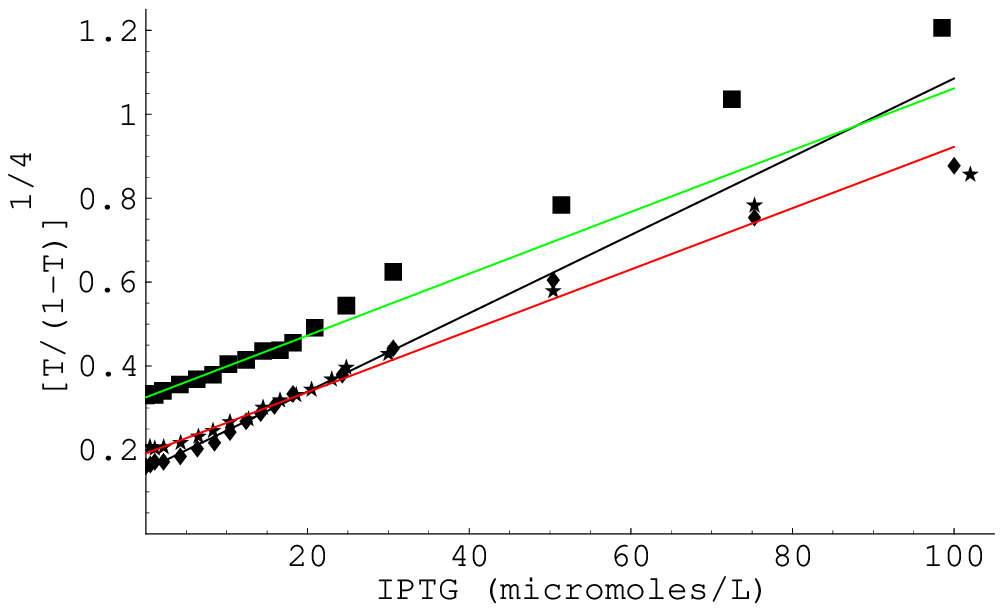}}\par\end{centering}

\caption{\label{f:OverathData}Analysis of the data for strains BB20 \emph{lac}$_{3}^{-}$
($\blacklozenge$), 2001c ($\bigstar$), and 15TAU \emph{lac}$_{2}^{-}$
($\blacksquare$)~\citep[Fig.~1]{Overath1968}. (a)~The $\left[T/(1-T)\right]^{1/2}$
vs.~$x$ plots are not straight lines. The slopes decrease significantly
at low inducer concentrations. (b)~The $\left[T/(1-T)\right]^{1/4}$
vs.~$x$ plots are linear at low inducer concentrations. The black,
red, and green lines are fits obtained from the data for IPTG concentrations
below 20~$\mu$M.}
\end{figure}

Analysis of the data confirms that DNA looping produces a qualitative
change in the kinetics, which cannot be captured by quantitative adjustment
of the parameters in eq.~(\ref{eq:Case1nu0}). If the data were consistent
with (\ref{eq:Case1nu0}), the $[T/(1-T{)]}^{1/2}$ vs.~$x$ plots
would be straight lines. However, construction of these plots for
three different strains of \emph{E. coli} yields not straight lines,
but curves with conspicuously small slopes at low inducer concentrations
(Fig.~\ref{f:OverathData}a).

\begin{table}

\caption{\label{t:Param}Parameter values of eq.~(\ref{eq:Case2T}) estimated
from the induction curves for 6 different strains of \emph{E. coli}.}

\begin{tabular}{|c|c|c|c|c|}
\hline
Strain&
$K_{x}^{-1}$ ($\mu$M)&
$\widehat{\alpha}_{1}$&
$\alpha_{1}$&
Reference\tabularnewline
\hline
\hline
BB20 \emph{lac}$_{3}^{-}$&
16.3&
1834&
62&
\citealp[Fig.~1]{Overath1968}\tabularnewline
\hline
2001c&
26.2&
741&
12&
\citealp[Fig.~1]{Overath1968}\tabularnewline
\hline
15 TAU \emph{lac}$_{2}^{-}$&
44.2&
89&
0&
\citealp[Fig.~1]{Overath1968}\tabularnewline
\hline
600Co$^{c}$$y_{1}^{-}$&
17.5&
13&
0&
\citealp[Fig.~1]{Overath1968}\tabularnewline
\hline
W3102$i^{t}$&
3.0&
66&
7&
\citeauthor[Fig.~1]{Gilbert1966}\tabularnewline
\hline
BMH8117 $\lambda$Ewt123&
10.9&
4921&
219&
\citealp[Fig.~1A]{Oehler2006}\tabularnewline
\hline
\end{tabular}
\end{table}

The reason for the nonlinearity of the $[T/(1-T)]^{1/2}$ vs.~$x$
plot becomes evident if eq.~(\ref{eq:Case2T}) is rewritten as\[
\frac{1}{T}-1=\frac{\widehat{\alpha}_{1}}{\left(1+\chi\right)^{4}}+\frac{\alpha_{1}}{\left(1+\chi\right)^{2}}.\]
Since $\widehat{\alpha}_{1}\sim50\alpha_{1}$ in wild-type \emph{lac},
the first term, which accounts for the repression due to looped unary
operons, dominates at sufficiently low inducer concentrations, $\chi\ll\sqrt{\widehat{\alpha}_{1}/\alpha_{1}}-1\approx6$.
At these low concentrations, $[T/(1-T)]^{1/4}$ vs.~$x$ plots should
be straight lines because\[
\left(\frac{T}{1-T}\right)^{1/4}\approx\frac{1}{\widehat{\alpha}_{1}^{1/4}}+\left(\frac{K_{x}}{\widehat{\alpha}_{1}^{1/4}}\right)x.\]
The experimental data for 3 different strains of \emph{E. coli} shows
that this is indeed the case (Fig.~\ref{f:OverathData}b). To be
sure, the $[T/(1-T)]^{1/2}$ vs.~$x$ plots are also straight lines
at sufficiently large inducer concentrations (Fig.~\ref{f:OverathData}a).
This is because when $\chi\gg\sqrt{\widehat{\alpha}_{1}/\alpha_{1}}-1$,
the non-looped unary states dominate, so that \[
\left(\frac{T}{1-T}\right)^{1/2}\approx\frac{1}{\alpha_{1}^{1/2}}+\left(\frac{K_{x}}{\alpha_{1}^{1/2}}\right)x.\]
However, neither plot can be linear over the entire range of inducer
concentrations.

\begin{figure}[t]
\noindent \begin{centering}\subfigure[]{\includegraphics[width=2.5in,keepaspectratio]{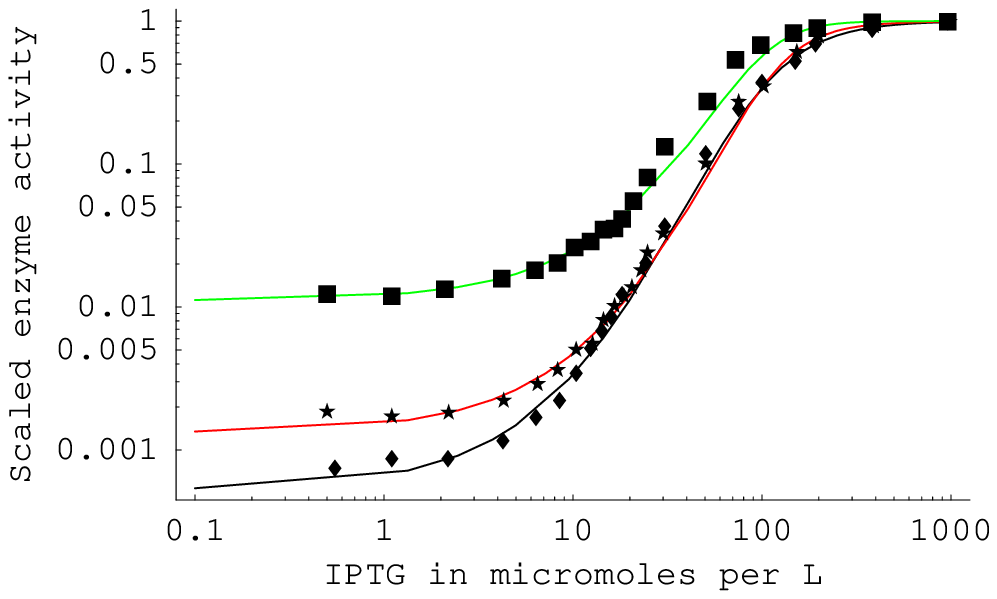}}\hspace*{0.3in}\subfigure[]{\includegraphics[width=2.5in,keepaspectratio]{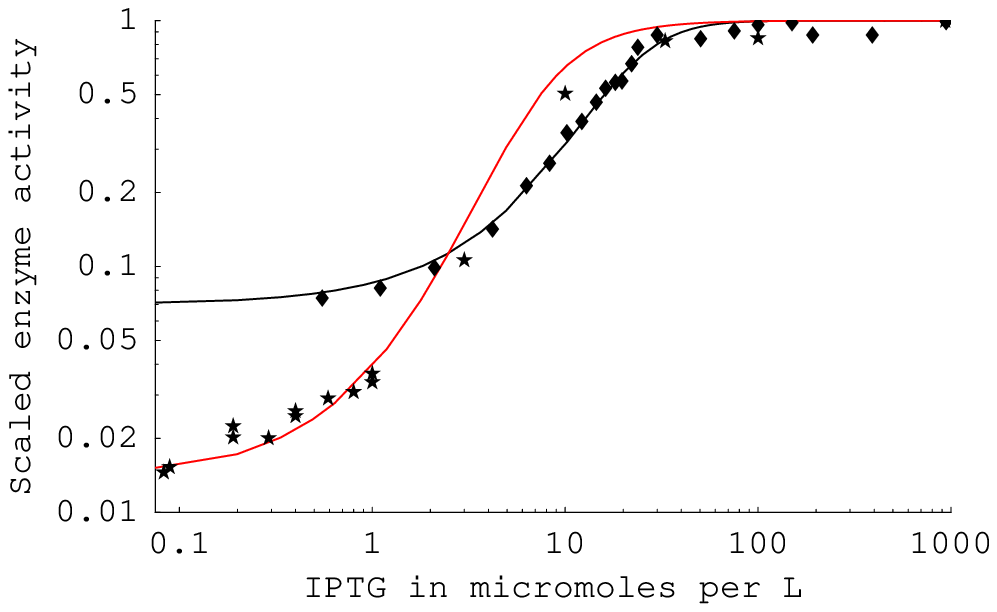}}\par\end{centering}

\caption{\label{f:OverathFits}Fits of the data from \citealp{Overath1968}
and \citealp{Gilbert1966} for: (a) BB20 \emph{lac}$_{3}^{-}$ ($\blacklozenge$),
2001c ($\bigstar$), and 15TAU \emph{lac}$_{2}^{-}$ ($\blacksquare$).
(b) Operator-constitutive strain 600Co$^{c}$$y_{1}^{-}$ ($\blacklozenge$),
and tight-binding strain W3102 ($\bigstar$). The data was fitted
with eq.~(\ref{eq:Case2T}) and the parameter values in Table~\ref{t:Param}.}
\end{figure}

Eq.~(\ref{eq:Case2T}) provides good fits to the experimental data
(Figs.~\ref{f:OehlerInductionData}c and~\ref{f:OverathFits}).
The parameter values for these fits, shown in Table~\ref{t:Param},
were estimated as follows. If sufficient data was available at low
inducer concentrations (Fig.~\ref{f:OverathFits}), $\widehat{\alpha}_{1}$
and $K_{x}$ were estimated from the slopes and intercepts of the
$[T/(1-T)]^{1/4}$ vs.~$x$ plots. The value of $\alpha_{1}$ was
then determined by one-parameter nonlinear regression of the data
(MATLAB, LSQNONLIN). If accurate data was not available at low concentrations
(Fig.~\ref{f:OehlerInductionData}c), all three parameter values
were obtained by nonlinear regression of the data.

\begin{figure}[t]
\noindent \begin{centering}\subfigure[]{\includegraphics[width=2.5in,height=1.5in]{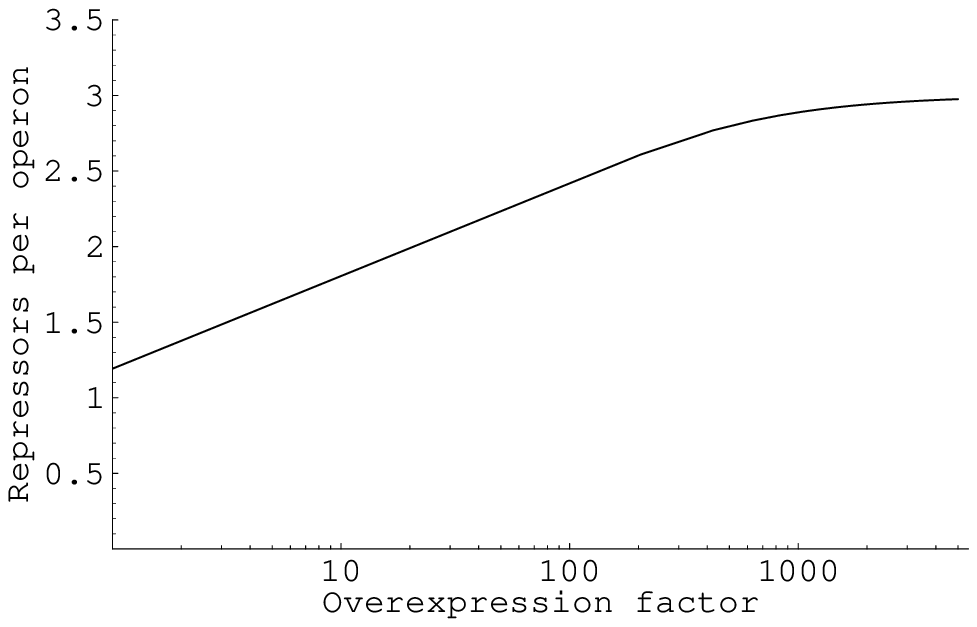}}\hspace*{0.3in}\subfigure[]{\includegraphics[width=2.5in,height=1.5in,keepaspectratio]{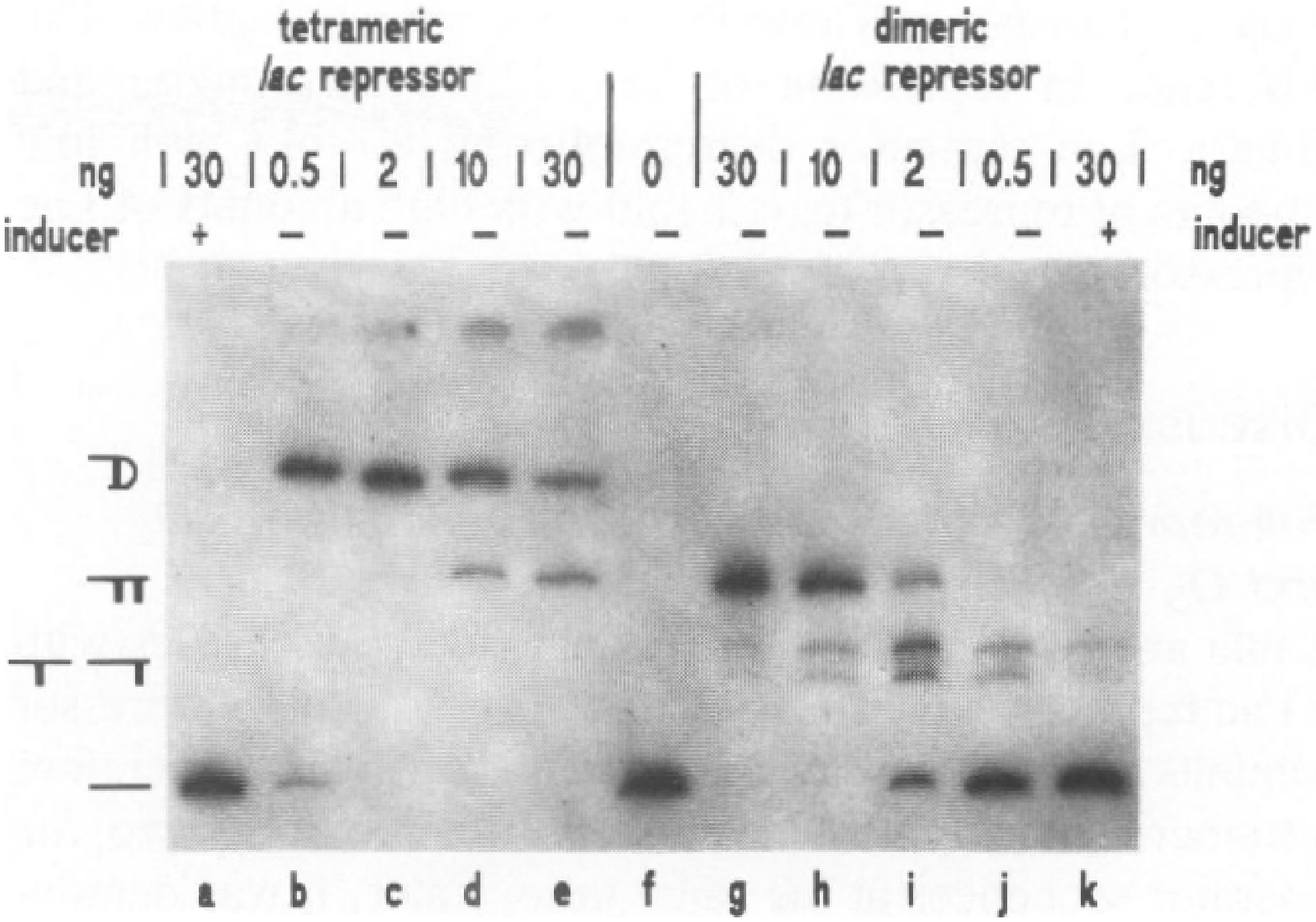}}\par\end{centering}

\caption{\label{f:TernaryFormation}The number of repressors per operon increases
with the fold-increase in repressor level relative to the wild-type
repressor level. (a)~The model prediction calculated from (\ref{eq:omega0})
assuming $\chi=0$ and $\alpha_{i},\widehat{\alpha}_{i}$ have wild-type
values. (b) When DNA fragments with two \emph{lac} operators are exposed
to increasing repressor levels (lanes b--e), the concentration of
binary non-looped fragments increases progressively~\citep[Fig.~4]{Oehler1990}.
The symbols on the left show the structures of the fragments (unary
looped at the top, followed by binary non-looped, unary non-looped,
and free fragments).}
\end{figure}

\begin{figure}[t]
\noindent \begin{centering}\subfigure[]{\includegraphics[width=1.8in,keepaspectratio]{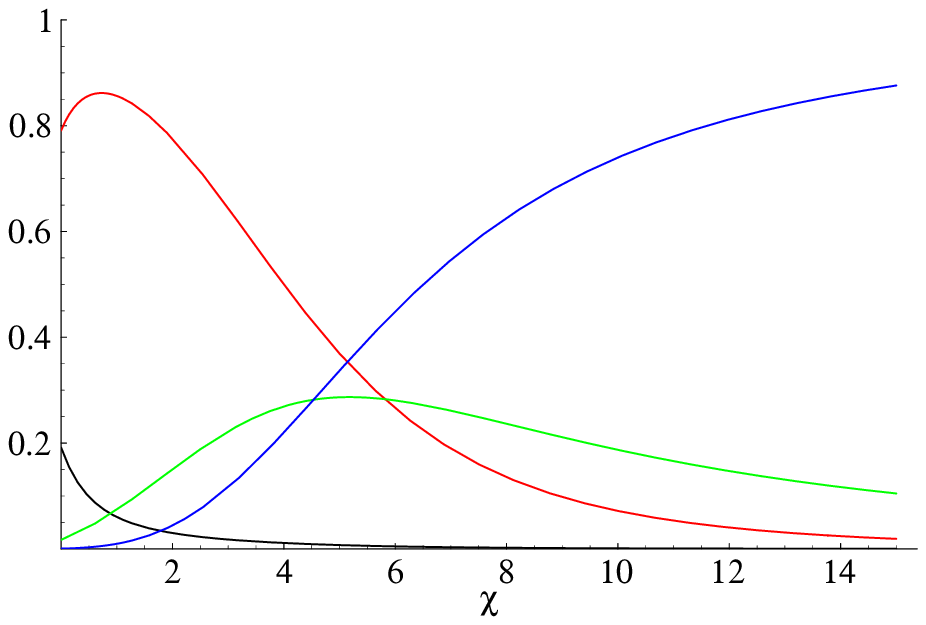}}\subfigure[]{\includegraphics[width=1.8in,keepaspectratio]{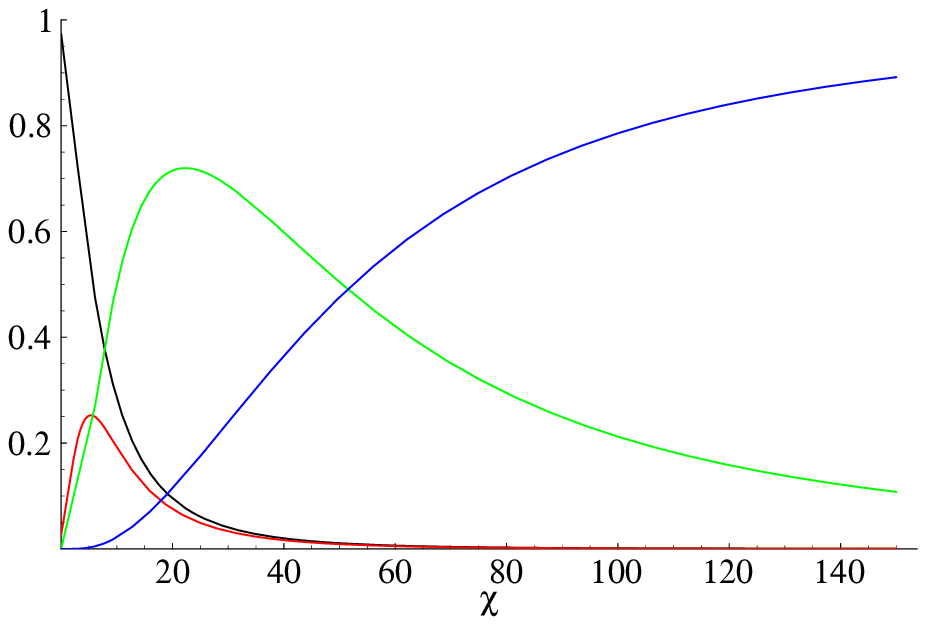}}\subfigure[]{\includegraphics[width=1.8in,keepaspectratio]{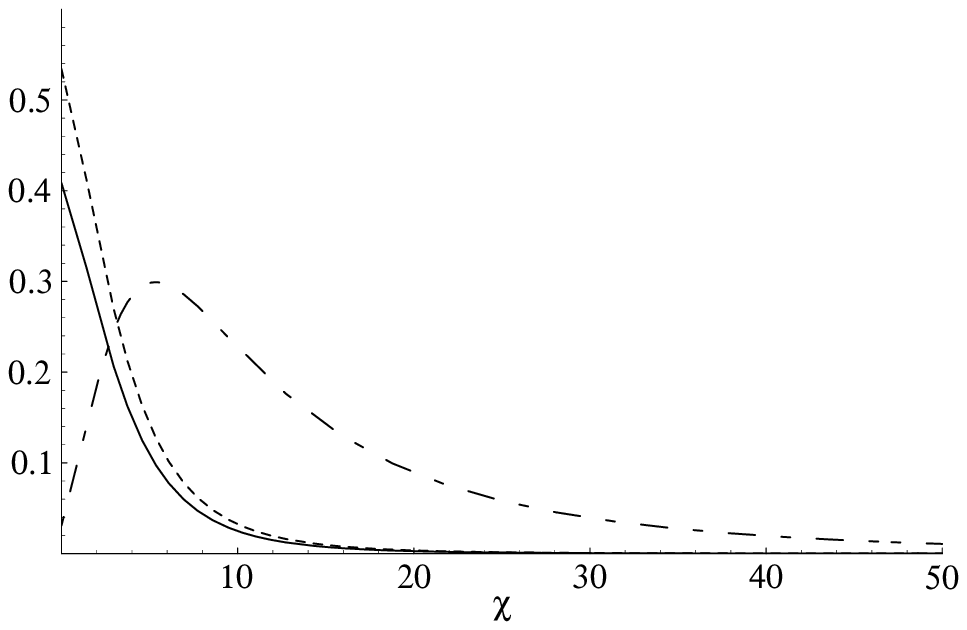}}\par\end{centering}

\caption{\label{f:Overexpressed}Distribution of the fractions of various
states as a function of the inducer concentration: (a)~Wild-type
cells. (b,c) Repressor-overpressed cells with 90-fold overexpression.
In (a, b), the black curve represents the fraction of binary and ternary
operons; the red, green, and blue curves represent the fractions of
looped unary, non-looped unary, and free operons, respectively. In
(c), the full, dashed, and long-dashed lines denote the fractions
of ternary, looped binary, and non-looped binary operons, respectively. }
\end{figure}

In wild-type cells, the binary and ternary operons were neglected
by appealing to (\ref{eq:phi})--(\ref{eq:phiHat}) and (\ref{eq:UnaryDominance}).
The latter relation is not valid for repressor-overexpressed cells.
This is because $\alpha_{j},\widehat{\alpha}_{j}$ are proportional
to $(r_{t})^{j}$. Hence, as the repressor level increases, $\alpha_{2},\widehat{\alpha}_{2},\alpha_{3}$
increase much faster than $\alpha_{1},\widehat{\alpha}_{1}$, and
at sufficiently large repressor levels,\begin{equation}
\alpha_{3}\gg\alpha_{2},\widehat{\alpha}_{2}\gg\alpha_{1},\widehat{\alpha}_{1}\gg1,\label{eq:alphaLargeRt}\end{equation}
i.e., almost all the operons are in the ternary state. Fig.~\ref{f:TernaryFormation}a
shows that in the absence of the inducer, $\Omega_{0}\approx1$ in
wild-type cells, but increases to $\sim$3 in cells containing $\sim$500
times the wild-type repressor levels. \emph{In vitro} data provides
direct evidence of this increase in $\Omega_{0}$. When DNA fragments,
containing two appropriately spaced \emph{lac} operators, are exposed
to increasing repressor levels, there is a perceptible increase in
the concentration of binary non-looped complexes (Fig.~\ref{f:TernaryFormation}b).
\emph{In vivo} data also suggests that $\Omega_{0}$ increases in
repressor-overexpressed cells. Oehler et al found similar repression
levels in two different strains of \emph{E. coli} containing high
levels (900 molecules per cell) of the wild-type tetrameric and mutant
dimeric repressor, respectively~\citep[Table~I]{Oehler1990}. They
argued that this is because at such high repressor levels, most of
the operons are in the ternary state. Since ternary operons cannot
form loops even in cells containing the tetrameric repressor, the
repression levels are similar in both cell types. More precisely,
(\ref{eq:alphaLargeRt}) and (\ref{eq:Dimer312}) imply that\[
\frac{\left.\mathcal{R}_{312}\right|_{\textnormal{dimer}}}{\left.\mathcal{R}_{312}\right|_{\textnormal{tetramer}}}=\frac{\left(1+\alpha_{1}/2+\alpha_{2}/4+\alpha_{3}/8\right)/\left(1+\kappa_{2}/2\right)}{\left(1+\alpha_{1}+\widehat{\alpha}_{1}+\alpha_{2}+\widehat{\alpha}_{2}+\alpha_{3}\right)/(1+\kappa_{2})}\approx\frac{1}{4}.\]
The experimentally observed value of this ratio is higher (0.5) possibly
because at such high tetrameric repressor levels, the repression is
too high to be measured accurately. The measured value of the repression
is, at best, a lower bound~\citep[Fig.~5]{Oehler1994}.

It is therefore clear that in repressor-overexpressed cells, binary
and ternary operons are dominant in the absence of the inducer. We
expect that they will remain dominant at sufficiently small inducer
concentrations. This becomes evident if we plot the \emph{fractions}
of various states of the operon as a function of the inducer concentration.
The fractions of non-looped and looped operons containing $i$ repressors
are given by\begin{align}
\theta_{i}(\chi) & \equiv\frac{\phi_{i}(\chi)}{1+\phi_{1}(\chi)+\widehat{\phi}_{1}(\chi)+\phi_{2}(\chi)+\widehat{\phi}_{2}(\chi)+\phi_{3}(\chi)},\label{eq:theta}\\
\widehat{\theta}_{i}(\chi) & \equiv\frac{\widehat{\phi}_{i}(\chi)}{1+\phi_{1}(\chi)+\widehat{\phi}_{1}(\chi)+\phi_{2}(\chi)+\widehat{\phi}_{2}(\chi)+\phi_{3}(\chi)}.\label{eq:thetaHat}\end{align}
The fraction of free operons, which is precisely $\nu$, is given
by (\ref{eq:nu0Phis}). In wild-type cells, the fraction of binary
and ternary operons, ($\theta_{2}+\widehat{\theta}_{2}+\theta_{3}$),
is small at all inducer concentrations (Fig~\ref{f:Overexpressed}a,
black curve). In repressor-overexpressed cells with 90-fold overexpression,
this fraction is dominant for all $\chi\lesssim5$ (Fig~\ref{f:Overexpressed}b,
black curve). If we plot the individual components, $\theta_{2},\widehat{\theta}_{2},\theta_{3}$,
of this fraction, it becomes clear that the ternary and binary looped
operons are dominant for $\chi\lesssim3$ (Fig~\ref{f:Overexpressed}c).
It follows that the kinetics of repressor-overexpressed cells cannot
be captured by eq.~(\ref{eq:Case2T}) --- it is necessary to use
the more general expression~(\ref{eq:Case3T}).

We tested the validity of the model by determining the extent to which
it could fit the induction curves for cells containing wild-type repressor
levels (Fig.~\ref{f:OverathFits}). The fits do not prove the validity
of the model because these induction curves show the variation of
only one of the model variables --- the fraction of free operons ---
as a function of the inducer concentration, . If the model is truly
valid, the fraction of every looped and non-looped species will vary
in a manner consistent with the model. It is therefore particularly
useful that these fractions follow simple scaling relations, which
are experimentally testable because each fraction migrates at a different
speed in polyacrylamide gel electrophoresis~(Fig.~\ref{f:TernaryFormation}b).
To see this, note that there are three distinct trends in Figs.~\ref{f:Overexpressed}b,c:
(a) The fraction of free operons increases monotonically, (b) the
fractions of ternary and looped binary operons decrease monotonically,
and (c) the fractions of the remaining three states of the operon
pass through a maximum. These trends follow immediately from the definitions
(\ref{eq:theta})--(\ref{eq:thetaHat}). They are similar to the concentration
profiles observed in series reactions ($A\rightarrow B\rightarrow\cdots$),
wherein as time progresses, the concentration of the first (resp.,
last) component decreases (resp., increases) monotonically, and the
concentrations of the intermediate components pass through a maximum.
In Figs.~\ref{f:Overexpressed}b,c, the inducer concentration plays
a role analogous to time: As $\chi$ increases, the ternary operons
are successively converted to binary, unary, and free operons. But
there is an important difference. Since $\widehat{\phi}_{2}$ and
$\phi_{3}$ decrease with $\chi$ at the same rate, the model predicts
that the ratio, $\widehat{\theta}_{2}/\theta_{3}$, has the same value,
$\widehat{\alpha}_{2}/\alpha_{3}$, at \emph{all} inducer concentrations.
Similarly, the ratio, $\widehat{\theta}_{1}/\theta_{2}$, must have
the same value, $\widehat{\alpha}_{1}/\alpha_{2}$, at all inducer
concentrations. These scaling relations were obtained by varying the
inducer concentrations at fixed repressor levels. If the repressor
levels are changed at fixed inducer levels, say, $\chi=0$ (Fig.~\ref{f:TernaryFormation}b),
the model predicts that $\widehat{\theta}_{i}/\theta_{i}$ will have
the same value, $\widehat{\alpha}_{i}/\alpha_{i}$, at all repressor
levels. Experimental tests of these scaling relations provide a stringent
test of the model. Furthermore, deviations from these scaling relations
may reveal the untenable assumptions of the model.

\section{Discussion}

\begin{figure}
\noindent \begin{centering}\includegraphics[width=2.8in,keepaspectratio]{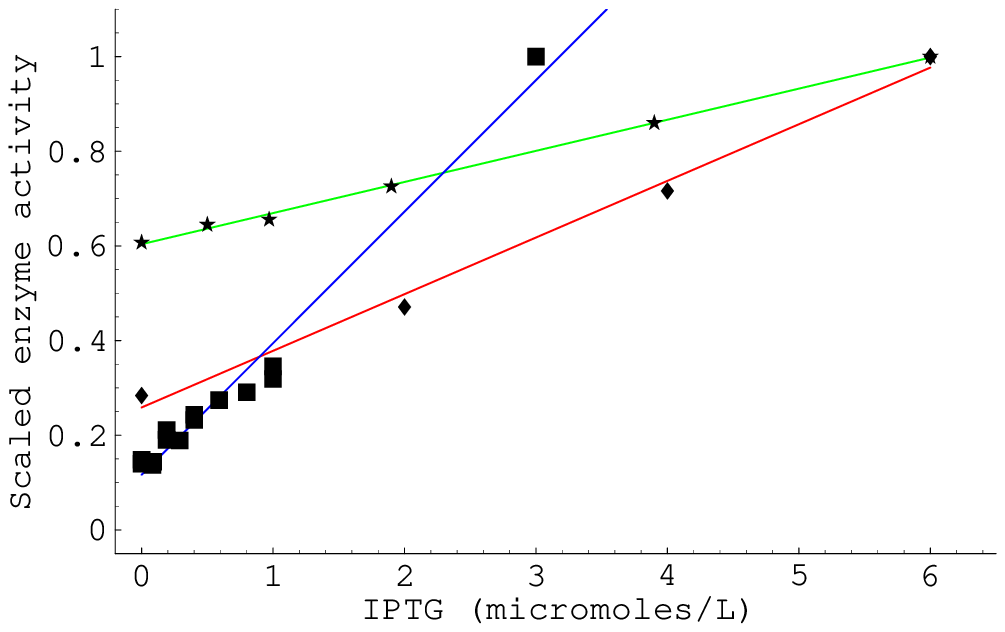}\par\end{centering}

\caption{\label{f:linearKinetics}The induction rate increases linearly at
small inducer concentrations~\citep{Gilbert1966,Oehler2006,Overath1968}.
The data corresponds to \emph{E. coli} BMH8117 $\lambda$Ewt100 ($\blacklozenge$),
which contains only the main operator, and \emph{E. coli} 15TAU \emph{lac}$_{2}^{-}$
($\bigstar$), W3102 ($\blacksquare$) which contain all three operators.}
\end{figure}

Given the above results, we can state the conditions under which the
kinetics of \emph{lac} induction can be described by eqs.~(\ref{eq:YagilGeneral})
and (\ref{eq:YagilSpecial}) of the Yagil \& Yagil model. If DNA looping
is weak or absent, both equations provide good approximations to the
kinetics, but (\ref{eq:YagilGeneral}) is valid at all inducer concentrations,
whereas (\ref{eq:YagilSpecial}) captures the kinetics only at sufficiently
large inducer concentrations. Indeed, the latter equation predicts
that the slope of the induction curve is zero at small inducer concentrations.
This is inconsistent with the data --- the induction curve increases
linearly at inducer concentrations as low as $\sim$0.5~$\mu$M,
regardless of the presence or absence of DNA looping (Fig.~\ref{f:linearKinetics}).

In the presence of DNA looping, the kinetics of wild-type cells are
more cooperative than the kinetics predicted by the Yagil \& Yagil
model, and this cooperativity becomes even more pronounced in repressor-overexpressed
cells. This result has important implications for the dynamics of
the \emph{lac} operon. As we show below, it suggests that repressor
overexpression can be used to induce bistability in systems that are
otherwise bistable.

\begin{figure}[t]
\noindent \begin{centering}\subfigure[]{\includegraphics[width=2.5in,keepaspectratio]{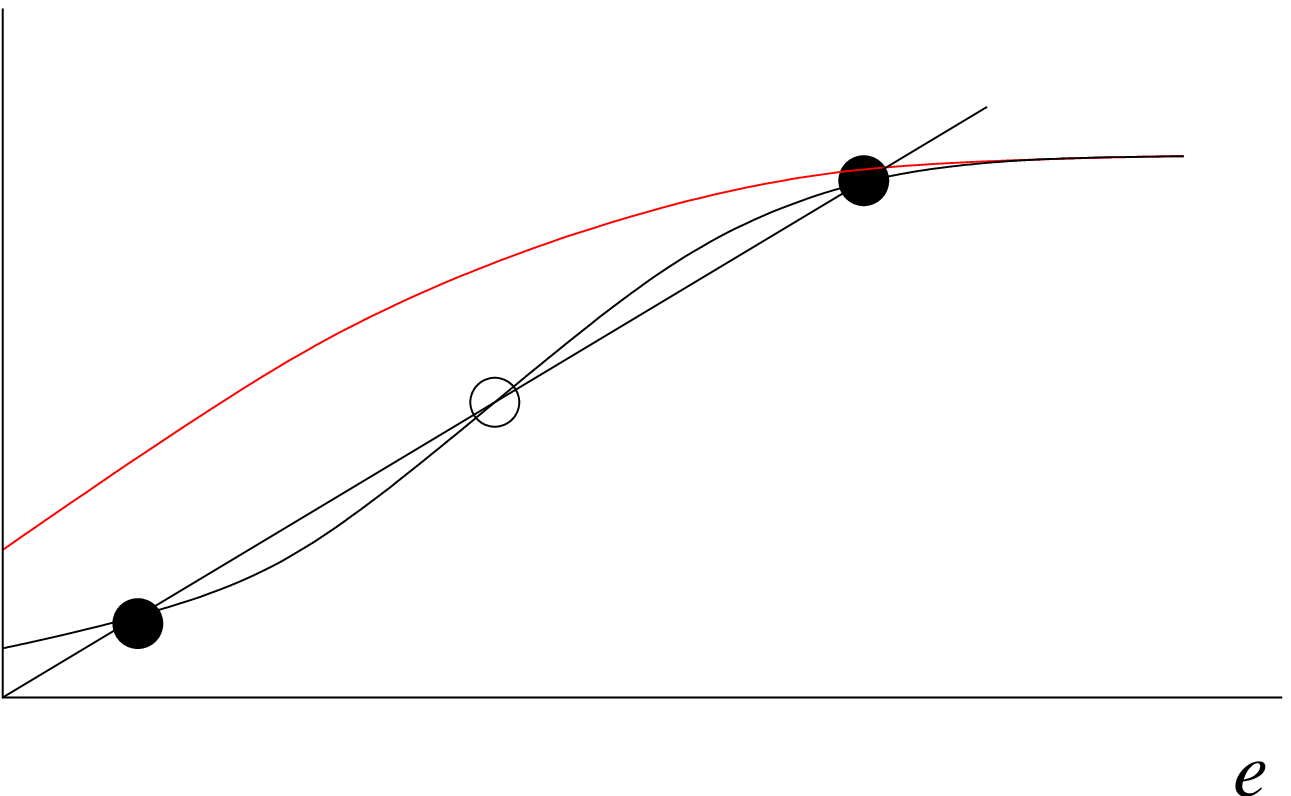}}\hspace*{0.2in}\subfigure[]{\includegraphics[width=2.2in,keepaspectratio]{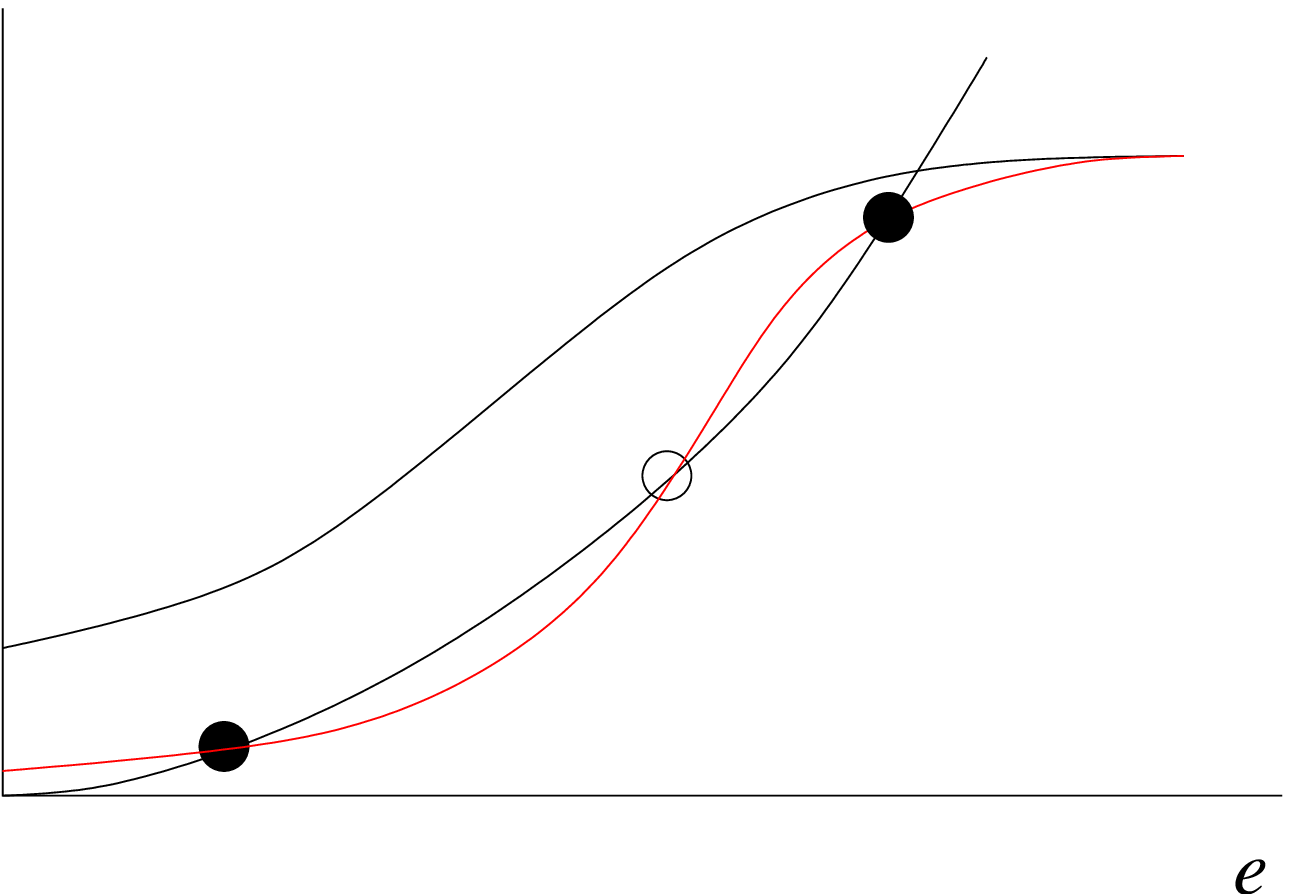}}\par\end{centering}

\caption{\label{f:Kinetics}Induction of bistability by repressor overexpression.
(a) In wild-type \emph{E. coli}, there is bistability during growth
on succinate + TMG (black curves). If the repressor levels are reduced,
the induction rate becomes hyperbolic (red curve), and bistability
disappears. (b) In wild-type \emph{E. coli}, there is no bistability
during growth on lactose (black curves). If the repressor is overexpressed,
bistability is induced because the induction rate becomes more cooperative
(red curve). }
\end{figure}

Molecular biologists have known for a long time that cooperativity
plays a central role in genetic switches~\citep[p.~28]{ptashne1}.
This was conclusively demonstrated by recent experiments with the
\emph{lac} operon. Ozbudak \emph{et al} inserted into the chromosome
of \emph{E. coli} MG 1655 a single copy of a \emph{lac} reporter gene
coding for green fluorescence protein. In these cells, the green fluorescence
intensity provides a measure of the instantaneous activity of the
\emph{lac} enzymes. They showed that when these cells were grown exponentially
on a medium containing succinate and the gratuitous inducer, TMG,
the enzyme activities displayed bistability. Futhermore, this bistability
could be captured by the steady states of the equation\[
\frac{de}{dt}=\frac{1+K_{x}^{2}x^{2}}{\alpha_{1}+1+K_{x}^{2}x^{2}}-r_{g}e,\; x\propto e\frac{s}{K_{s}+s}\]
where $e$ and $s$ denote the \emph{lac} permease activity and extracellular
TMG concentration, respectively; $r_{g}$ denotes the specific growth
rate on succinate; and the inducer concentration, $x$, is assumed
to be proportional to the TMG uptake rate.%
\footnote{The repression of the \emph{lac} reporter gene used in this study
was only 170. This is partly because the reporter gene lacks $O_{2}$.
However, the $O_{1},$$O_{3}$ interaction is also somewhat attenuated
because $O_{1},O_{3}$-containing cells yield a repression of 440~(Table~\ref{t:RepressionData}).
Given the weak DNA looping, it is conceivable that eq.~(\ref{eq:YagilSpecial})
approximates the induction kinetics.%
} Bistability occurs precisely because the induction rate, which increases
as $e^{2}$, is more cooperative than the dilution rate, which is
proportional to $e$ (Fig.~\ref{f:Kinetics}a, black curves). Indeed,
if the repressor level is decreased by {}``titrating'' the repressor
with the \emph{lac} operator, the induction curve loses its cooperativity
--- it becomes hyperbolic (Fig.~\ref{f:Kinetics}a, red curve), and
the bistability disappears.

The above example shows that bistability can be abolished by decreasing
the repressor level, and hence, the cooperativity of the induction
curve. It is therefore conceivable that bistability can be imposed
upon monostable systems by increasing the repressor level. Ozbudak
et al observed that their system exhibited no bistability if the cells
were grown on lactose, rather than succinate + TMG~\citep{ozbudak04}.
One hypothesis for explaining the absence of bistability is as follows~\citep{Narang2007c}.
During growth on succinate + TMG, the specific growth rate is independent
of the \emph{lac} permease activity. In sharp contrast, during growth
on lactose, the specific growth rate is proportional to the specific
lactose uptake rate, i.e., $r_{g}\propto es/(K_{s}+s)$, where $s$
now represents the concentration of extracellular lactose. The dilution
rate is therefore as cooperative as the induction rate (both rates
increase as $e^{2}$), and bistability is impossible (Fig.~\ref{f:Kinetics}b,
black curves). In such systems, bistability can be induced by overexpressing
the repressor because the induction rate then increases as $e^{4}$
or $e^{6}$, which is significantly more cooperative than the dilution
rate (Fig.~\ref{f:Kinetics}b, red curve). Thus, the increase in
cooperativity generated by high repressor levels can be exploited
to impose bistability upon systems that otherwise show little propensity
for switch-like behavior. This may be useful in synthetic biology,
which is concerned, among other things, with the development of genetic
switches.

\section{Conclusions}

We formulated a model for the kinetics of \emph{lac} induction which
takes due account of the tetrameric structure of the repressor, the
existence of the auxiliary operators, and the attendant DNA looping.
Analysis of the model shows that:

\begin{enumerate}
\item In the absence of DNA looping, the kinetics are given by eq.~(\ref{eq:Case1nu0}),
which is formally similar to the Yagil \& Yagil model. In the presence
of DNA looping, the kinetics are significantly more cooperative.
\item In wild-type cells, no more than one repressor binds to an operon,
and the kinetics are given by eq.~(\ref{eq:Case2nu0}), which depends
on powers of $x$ as high as $x^{4}$. The cooperativity increases
markedly because the concentration of looped repressor-operator complexes
decreases with the inducer concentration at a rate much faster than
the corresponding rate for non-looped complexes.
\item If the repressor is overexpressed in wild-type cells, multiple repressors
are bound to most of the operons, and the kinetics are given by eq.~(\ref{eq:Case3nu0}),
which depends on powers of $x$ up to $x^{6}$. The cooperativity
is enhanced even further because multi-repressor operons are more
sensitive to the inducer concentrations than operons with only one
repressor.
\item The model provides good fits to the induction curves for 4 different
strains of \emph{E. coli}. We also show that if the model is correct,
the relative concentrations of certain looped and non-looped species
must remain the same at all inducer (or repressor) concentrations.
These scaling relations, which lie at the heart of the model, can
be rigorously tested by gel electrophoresis.
\end{enumerate}
These results should be useful in analyzing kinetic data for induction
of operons involving DNA looping, and in formulating dynamic models
for induction of such operons.

\begin{ack}
I thank the anonymous reviewers for their valuable comments. This
research was supported in part with funds from the National Science
Foundation under contract NSF DMS-0517954.
\end{ack}
\appendix

\section{\label{a:MutantDimers}Induction kinetics and repression in cells
containing mutant dimers}

Equations (\ref{eq:rho})--(\ref{eq:nu}) were derived for cells containing
the tetrameric repressor. If the cells contain mutant dimers that
can bind to the operator but do not tetramerize, the corresponding
equations are \begin{alignat}{1}
\rho\left(1+\chi\right)^{2}+\omega\nu\left[\rho\bar{\alpha}_{1}+2\rho^{2}\bar{\alpha}_{2}+3\rho^{3}\bar{\alpha}_{3}\right] & =1,\label{eq:rhoDimer}\\
\nu\left[1+\rho\bar{\alpha}_{1}+\rho^{2}\bar{\alpha}_{2}+\rho^{3}\bar{\alpha}_{3}\right] & =1.\label{eq:nuDimer}\end{alignat}
where $\rho$ now denotes the fraction of free mutant dimers. These
equations differ from eqs.~(\ref{eq:rho})--(\ref{eq:nu}) in three
ways: (a) The parameters, $\bar{\alpha}_{i}$, satisfy the relations,
$\bar{\alpha}_{1}=\alpha_{1}/2$, $\bar{\alpha}_{2}=\alpha_{2}/4$,
$\bar{\alpha}_{3}=\alpha_{3}/8$, since the association constants
for dimer-operator binding are half of the corresponding association
constants for tetramer-operator binding. (b) The first term of eq.~(\ref{eq:rhoDimer})
depends on $(1+\chi)^{2}$, rather than $(1+\chi)^{4}$, because mutant
dimers have only two inducer-binding sites. (c) The terms in square
brackets do not depend on the inducer concentrations because inducer-bound
mutant dimers cannot bind to the operator. The latter also implies
that the transcription rate is proportional to $T=\nu\left(1+\bar{\kappa}_{2}\rho\right),$
$\bar{\kappa}_{2}=\kappa_{2}/2$, provided $d=0$.

The zeroth-order solution is \begin{align*}
\rho_{0} & =\frac{1}{\left(1+\chi\right)^{2}},\\
\nu_{0} & =\frac{1}{1+\rho_{0}\bar{\alpha}_{1}+\rho_{0}^{2}\bar{\alpha}_{2}+\rho_{0}^{3}\bar{\alpha}_{3}},\end{align*}
which implies that\[
T=\frac{1+\bar{\kappa}_{2}/\left(1+\chi\right)^{2}}{1+\bar{\alpha}_{1}/\left(1+\chi\right)^{2}+\bar{\alpha}_{2}/\left(1+\chi\right)^{4}+\bar{\alpha}_{3}/\left(1+\chi\right)^{6}}.\]
Although these kinetics can be highly cooperative, the parameter values
for cells containing wild-type repressor levels are such that the
corresponding kinetics are formally similar to eq.~(\ref{eq:Case1T}).
Fig.~\ref{f:OehlerInductionData}b shows that this equation provides
a good fit to the induction curve of cells containing mutant dimers.

\begin{figure}
\noindent \begin{centering}\subfigure[]{\includegraphics[width=1.8in]{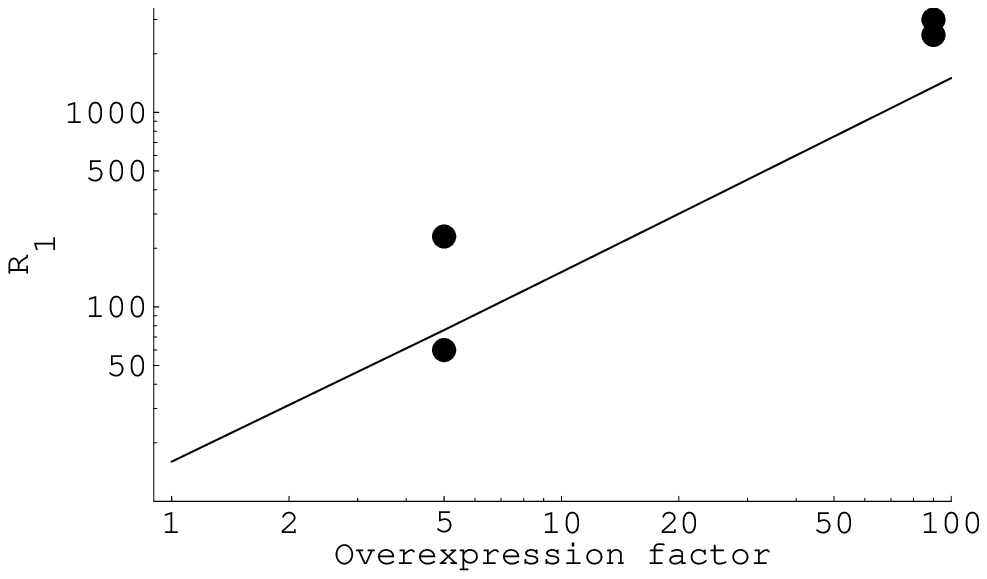}}\hspace*{0.0in}\subfigure[]{\includegraphics[width=1.8in]{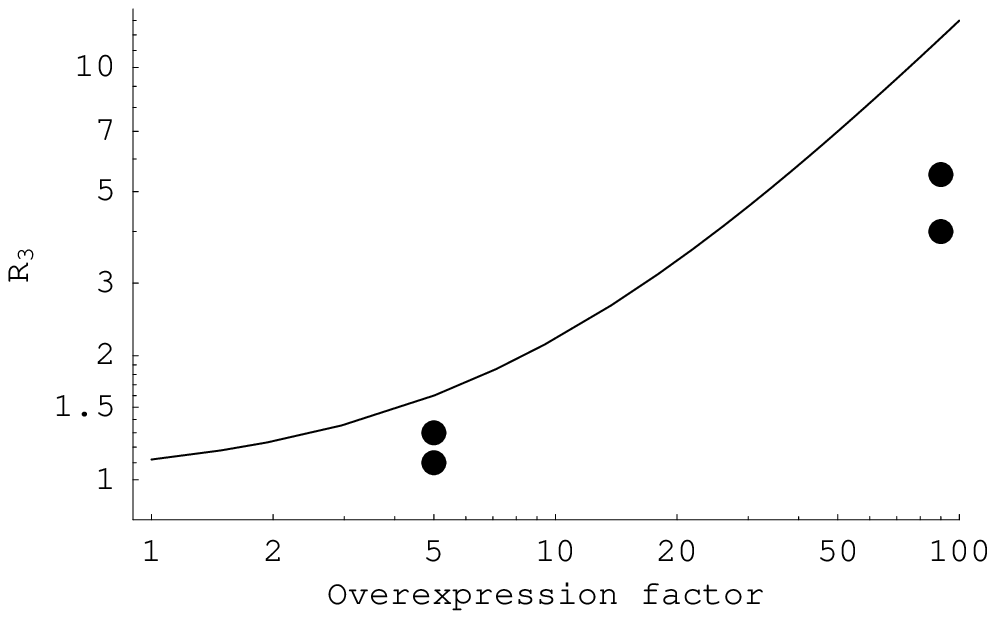}}\hspace*{0.0in}\subfigure[]{\includegraphics[width=1.8in]{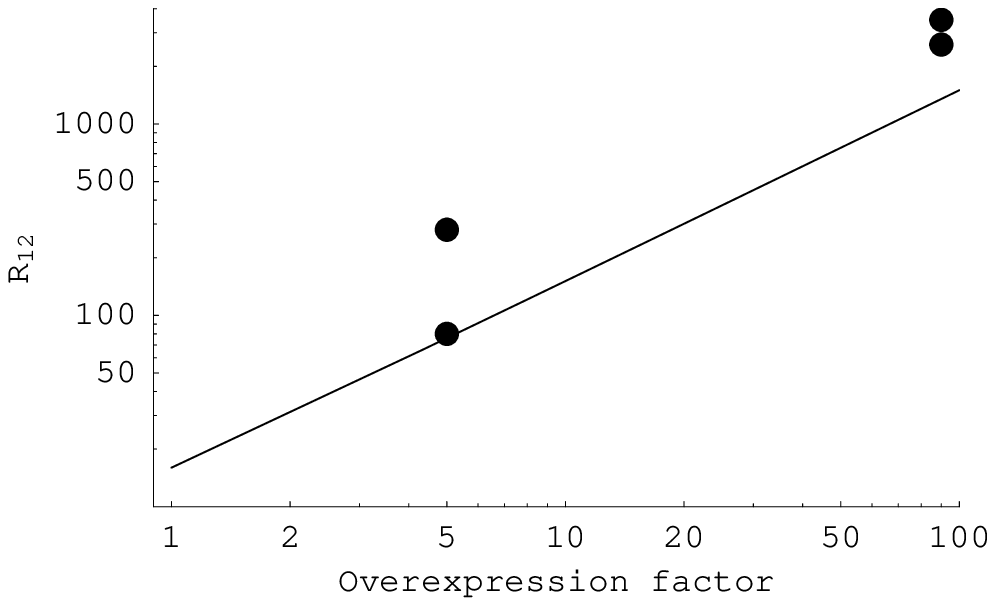}}\par\end{centering}

\noindent \begin{centering}\subfigure[]{\includegraphics[width=1.8in]{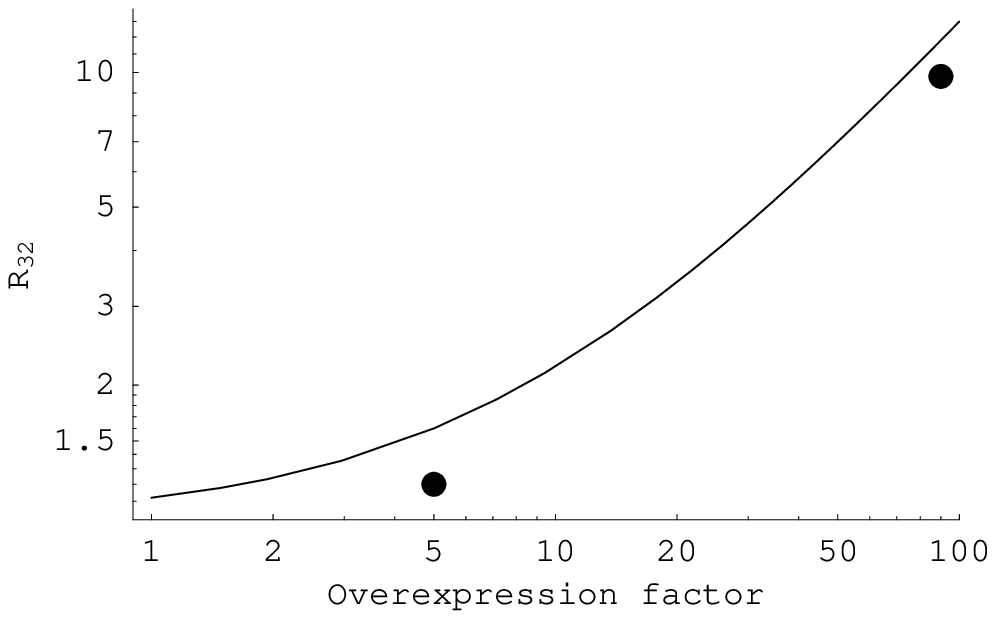}}\hspace*{0.0in}\subfigure[]{\includegraphics[width=1.8in]{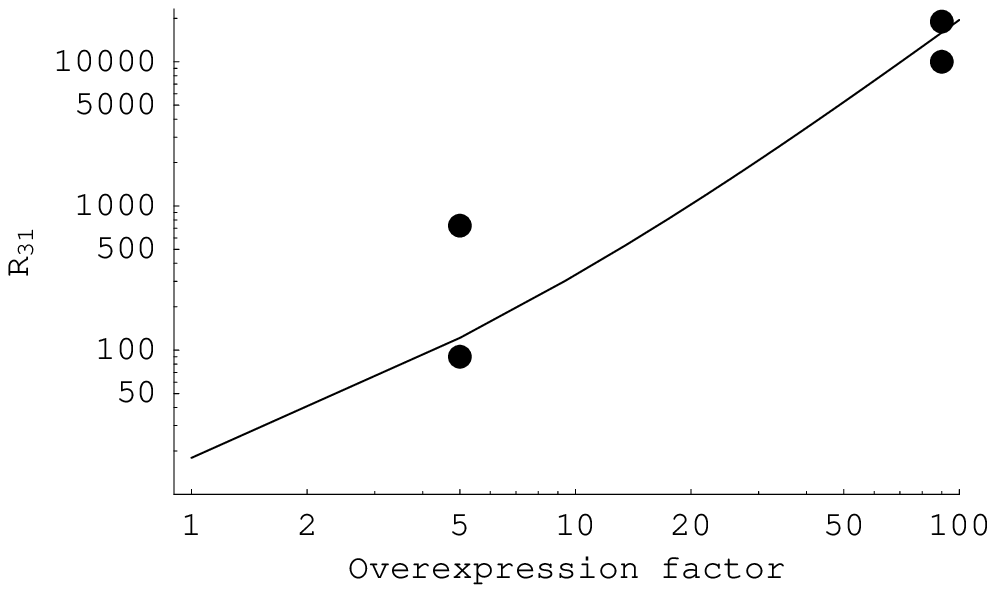}}\hspace*{0.0in}\subfigure[]{\includegraphics[width=1.8in]{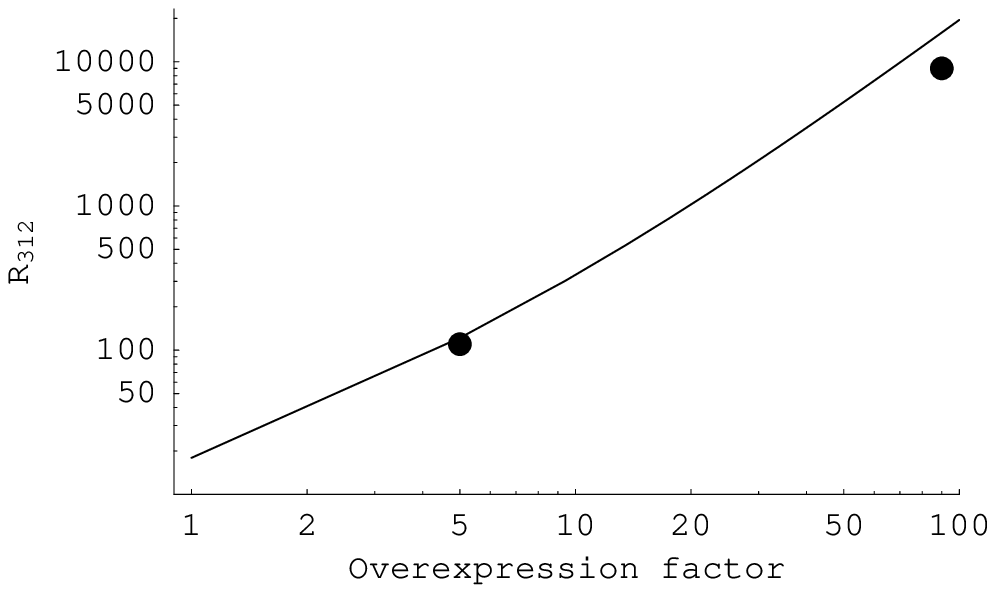}}\par\end{centering}

\caption{\label{f:DimerRepression}The model predicts the repression in cells
containing mutant dimers (data from \citealp[Table I]{Oehler1990},
and \citealp[Figs~4--5]{Oehler1994}). The full lines show the model
predictions, calculated from eqs.~(\ref{eq:Dimer312})--(\ref{eq:Dimer3})
with the wild-type parameter values, $\kappa_{1}=30$, $\kappa_{2}=0.38$,
$\kappa_{3}=0.24$, determined in Section~\ref{s:Results}.}
\end{figure}

To see this, observe that the values of $\alpha_{1}$, $\alpha_{2}$,
$\alpha_{3}$ for cells containing wild-type levels of tetrameric
repressor imply that $\bar{\alpha}_{1}=15.5$, $\bar{\alpha}_{2}=5$,
$\bar{\alpha}_{3}=0.25$. Since $\bar{\alpha}_{2},\bar{\alpha}_{3}$
are small compared to $\bar{\alpha}_{1}$, \[
\frac{\bar{\alpha}_{2}}{\left(1+\chi\right)^{4}},\frac{\bar{\alpha}_{3}}{\left(1+\chi\right)^{6}}\ll\frac{\bar{\alpha}_{1}}{\left(1+\chi\right)^{2}}\]
 for all but a negligibly small range of inducer concentrations. The
induction kinetics are therefore formally identical to eq.~(\ref{eq:Case1T}).

The repression in cells containing all three operators is\begin{equation}
\mathcal{R}_{312}=\frac{1+\bar{\alpha}_{1}+\bar{\alpha}_{2}+\bar{\alpha}_{3}}{1+\bar{\kappa}_{2}},\label{eq:Dimer312}\end{equation}
which implies that\begin{align}
\mathcal{R}_{1} & =1+\bar{\kappa}_{1}, & \mathcal{R}_{2} & =1,\label{eq:Dimer1}\\
\mathcal{R}_{3} & =1+\bar{\kappa}_{3}, & \mathcal{R}_{12} & =\frac{1+\bar{\kappa}_{1}+\bar{\kappa}_{2}+\bar{\kappa}_{1}\bar{\kappa}_{2}}{1+\bar{\kappa}_{2}},\label{eq:Dimer2}\\
\mathcal{R}_{32} & =\frac{1+\bar{\kappa}_{2}+\bar{\kappa}_{3}+\bar{\kappa}_{2}\bar{\kappa}_{3}}{1+\bar{\kappa}_{2}}, & \mathcal{R}_{31} & =1+\bar{\kappa}_{1}+\bar{\kappa}_{3}+\bar{\kappa}_{1}\bar{\kappa}_{3},\label{eq:Dimer3}\end{align}
where $\bar{\kappa}_{i}=\kappa_{i}/2$. Fig.~\ref{f:DimerRepression}
shows the repression predicted by these expressions, assuming that
$\kappa_{1},\kappa_{2},\kappa_{3}$ have the values estimated in Section~\ref{s:Results}
from the data for cells containing the tetrameric repressor. The good
agreement with the repression data for cells containing mutant dimers
suggests that the model and the parameter values are plausible.

\section{\label{a:MolecularParameters}Solution of eqs.~(\ref{eq:rho})--(\ref{eq:nu})
by regular perturbation}

We wish to solve the equations \begin{alignat}{1}
\rho\left(1+\chi\right)^{4}+\omega\nu(\rho f_{1}+2\rho^{2}f_{2}+3\rho^{3}f_{3}) & =1,\label{eq:rhoApp}\\
\nu(1+\rho f_{1}+\rho^{2}f_{2}+\rho^{3}f_{3}) & =1,\label{eq:nuApp}\end{alignat}
for small $\omega$. To this end, assume that the solutions have the
form\begin{align}
\rho & =\rho_{0}+\omega\rho_{1}+O(\omega^{2}),\label{eq:AssumedSoln1}\\
\nu & =\nu_{0}+\omega\nu_{1}+O(\omega^{2}).\label{eq:AssumedSoln2}\end{align}
Substituting these solutions in (\ref{eq:rhoApp})--(\ref{eq:nuApp}),
and collecting terms with like powers of $\omega$ yields\begin{align*}
\left[\rho_{0}\left(1+\chi\right)^{4}-1\right]+\omega\left[\rho_{1}\left(1+\chi\right)^{4}+\nu_{0}\left(\rho_{0}f_{1}+2\rho_{0}^{2}f_{2}+3\rho_{0}^{3}f_{3}\right)\right]+\ldots & =0,\\
\left[\nu_{0}\left(1+\rho_{0}f_{1}+\rho_{0}^{2}f_{2}+\rho_{0}^{3}f_{3}\right)-1\right]+\omega\left[\nu_{0}\rho_{1}\left(f_{1}+2\rho_{0}f_{2}+3\rho_{0}^{2}f_{3}\right)+\right.\\
+\left.\nu_{1}\left(1+\rho_{0}f_{1}+\rho_{0}^{2}f_{2}+\rho_{0}^{3}f_{3}\right)\right]+\ldots & =0.\end{align*}
It follows that\begin{align*}
\rho_{0} & =\frac{1}{\left(1+\chi\right)^{4}},\\
\nu_{0} & =\frac{1}{1+\rho_{0}f_{1}+\rho_{0}^{2}f_{2}+\rho_{0}^{3}f_{3}},\\
\rho_{1} & =-\frac{\nu_{0}}{\left(1+\chi\right)^{4}}\left(\rho_{0}f_{1}+2\rho_{0}^{2}f_{2}+3\rho_{0}^{3}f_{3}\right),\\
\nu_{1} & =-\nu_{0}\rho_{1}\frac{f_{1}+2\rho_{0}f_{2}+3\rho_{0}^{2}f_{3}}{1+\rho_{0}f_{1}+\rho_{0}^{2}f_{2}+\rho_{0}^{3}f_{3}}.\end{align*}
If we define\begin{equation}
\Omega_{0}\equiv\frac{\rho_{0}f_{1}+2\rho_{0}^{2}f_{2}+3\rho_{0}^{2}f_{3}}{1+\rho_{0}f_{1}+\rho_{0}^{2}f_{2}+\rho_{0}^{3}f_{3}},\label{eq:Omega0defn}\end{equation}
$\rho_{1}$ and $\nu_{1}$ can be written as \[
\rho_{1}=-\rho_{0}\Omega_{0},\;\nu_{1}=\nu_{0}\Omega_{0}^{2}.\]
Substituting these expressions in (\ref{eq:AssumedSoln1})--(\ref{eq:AssumedSoln2})
yields \begin{align}
\rho & =\rho_{0}\left(1-\omega\Omega_{0}\right)+O(\omega^{2}),\label{eq:FirstOrder1}\\
\nu & =\nu_{0}\left(1+\omega\Omega_{0}^{2}\right)+O(\omega^{2}).\label{eq:FirstOrder2}\end{align}
These are the first-order solutions for the general model.

\begin{figure}
\noindent \begin{centering}\includegraphics[width=3.5in,keepaspectratio]{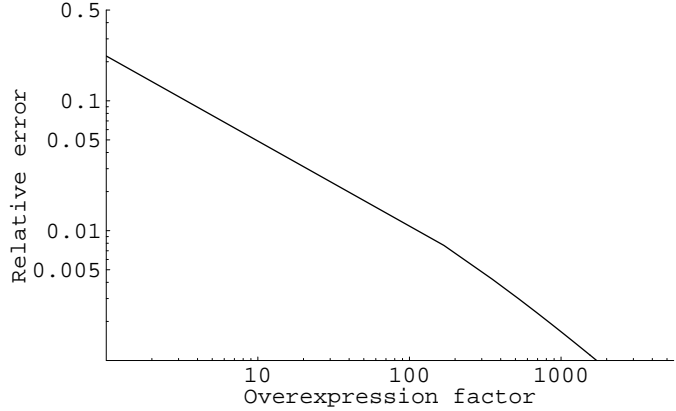}\par\end{centering}

\caption{\label{f:errorPlot}The relative error for $\nu_{0}$ does not exceed
$\sim$20\%. The relative error was calculated assuming $\chi=0$,
$\omega=0.2$, and $\alpha_{i},\widehat{\alpha}_{i}$ have wild-type
values.}
\end{figure}

The parameter $\Omega_{0}$ approximates the average number of repressors
bound to an operon because (\ref{eq:Omega0defn}) can be rewritten
as\[
\Omega_{0}=\theta_{1,t}+2\theta_{2,t}+3\theta_{3,t},\]
where \[
\theta_{i,t}\equiv\frac{\rho_{0}^{i}f_{i}}{1+\rho_{0}f_{1}+\rho_{0}^{2}f_{2}+\rho_{0}^{3}f_{3}},\; i=1,2,3.\]
is the fraction of operons containing $i$ repressors. It follows
that $\Omega_{0}$ must lie between 0 and 3. In the absence of the
inducer, $\Omega_{0}$ increases with repressor overexpression from
$\sim$1 to 3 (Fig.~\ref{f:TernaryFormation}). However, the relative
error for $\nu_{0}$ does not exceed $\sim$20\% (Fig.~\ref{f:errorPlot}).

In the absence of the auxiliary operators, $f_{2}=f_{3}=0$. In this
case \[
\nu_{0}=\frac{1}{1+\rho_{0}f_{1}},\;\Omega_{0}=\frac{\rho_{0}f_{1}}{1+\rho_{0}f_{1}}\Rightarrow\Omega_{0}=1-\nu_{0}.\]
Substituting this relation in (\ref{eq:FirstOrder1})--(\ref{eq:FirstOrder2})
yields\begin{align*}
\rho & =\rho_{0}\left[1-\omega\left(1-\nu_{0}\right)\right]+O(\omega^{2}),\\
\nu & =\nu_{0}\left[1+\omega\left(1-\nu_{0}\right)^{2}\right]+O(\omega^{2}).\end{align*}

\bibliographystyle{C:/texmf/bibtex/bst/elsevier/elsart-harv}

\begin{thebibliography}{19}
\expandafter\ifx\csname
natexlab\endcsname\relax\def\natexlab#1{#1}\fi
\expandafter\ifx\csname url\endcsname\relax
  \def\url#1{\texttt{#1}}\fi
\expandafter\ifx\csname urlprefix\endcsname\relax\def\urlprefix{URL
}\fi

\bibitem[{Ackers et~al.(1982)Ackers, Johnson, and Shea}]{Ackers1982}
Ackers, G.~K., Johnson, A.~D., Shea, M.~A., Feb 1982. Quantitative
model for
  gene regulation by lambda phage repressor. Proc Natl Acad Sci U S A 79~(4),
  1129--1133.

\bibitem[{Barry and Matthews(1999)}]{Barry1999}
Barry, J.~K., Matthews, K.~S., May 1999. Thermodynamic analysis of
unfolding
  and dissociation in lactose repressor protein. Biochemistry 38~(20),
  6520--6528.

\bibitem[{Carroll et~al.(2005)Carroll, Grenier, and Weatherbee}]{Carroll}
Carroll, S.~B., Grenier, J.~K., Weatherbee, S.~D., 2005. From DNA to
Diversity:
  Molecular Genetics and the Evolution of Animal Design, 2nd Edition. Blackwell
  Publishing.

\bibitem[{Chung and Stephanopoulos(1996)}]{chung96}
Chung, J.~D., Stephanopoulos, G., 1996. On physiological
multiplicity and
  population heterogeneity of biological systems. Chem.\ Eng.\ Sc. 51,
  1509--1521.

\bibitem[{Gilbert and Müller-Hill(1966)}]{Gilbert1966}
Gilbert, W., Müller-Hill, B., Dec 1966. Isolation of the
\textit{lac}
  repressor. Proc Natl Acad Sci U S A 56~(6), 1891--1898.

\bibitem[{Herzenberg(1959)}]{Herzenberg1959}
Herzenberg, L.~A., Feb 1959. Studies on the induction of
$\beta$-galactosidase
  in a cryptic strain of \textit{{E}scherichia coli}. Biochim Biophys Acta
  31~(2), 525--538.

\bibitem[{Laurent et~al.(2005)Laurent, Charvin, and
  Guespin-Michel}]{Laurent2005}
Laurent, M., Charvin, G., Guespin-Michel, J., Dec 2005. Bistability
and
  hysteresis in epigenetic regulation of the lactose operon. {S}ince
  {D}elbrück, a long series of ignored models. Cell Mol Biol (Noisy-le-grand)
  51~(7), 583--594.

\bibitem[{Lewis(2005)}]{Lewis2005}
Lewis, M., Jun 2005. The lac repressor. C R Biol 328~(6), 521--548.

\bibitem[{M\"uller-Hill(1996)}]{Muller-Hill}
M\"uller-Hill, B., 1996. The \textit{lac} operon, 1st Edition. de
Gruyter,
  Berlin.

\bibitem[{Narang and Pilyugin(2006)}]{Narang2007c}
Narang, A., Pilyugin, S.~S., December 2006. Why does the
\textit{lac} exhibit
  no bistability during growth of \textit{Escherichia coli} on lactose or
  lactose + glucose?, submitted, Bull. Math. Biol.

\bibitem[{Oehler et~al.(2006)Oehler, Alberti, and Müller-Hill}]{Oehler2006}
Oehler, S., Alberti, S., Müller-Hill, B., 2006. {I}nduction of the
\textit{lac}
  promoter in the absence of {DNA} loops and the stoichiometry of induction.
  Nucleic Acids Res 34~(2), 606--612.

\bibitem[{Oehler et~al.(1994)Oehler, Amouyal, Kolkhof, von Wilcken-Bergmann,
  and Müller-Hill}]{Oehler1994}
Oehler, S., Amouyal, M., Kolkhof, P., von Wilcken-Bergmann, B.,
Müller-Hill,
  B., Jul 1994. Quality and position of the three \textit{lac} operators of
  \textit{{E}. coli} define efficiency of repression. EMBO J 13~(14),
  3348--3355.

\bibitem[{Oehler et~al.(1990)Oehler, Eismann, Krämer, and
  Müller-Hill}]{Oehler1990}
Oehler, S., Eismann, E.~R., Krämer, H., Müller-Hill, B., Apr 1990.
The three
  operators of the \textit{lac} operon cooperate in repression. EMBO J 9~(4),
  973--979.

\bibitem[{Overath(1968)}]{Overath1968}
Overath, P., 1968. Control of basal level activity of
$\beta$-galactosidase in
  \textit{{E}scherichia coli}. Mol Gen Genet 101~(2), 155--165.

\bibitem[{Ozbudak et~al.(2004)Ozbudak, Thattai, Lim, Shraiman, and van
  Oudenaarden}]{ozbudak04}
Ozbudak, E.~M., Thattai, M., Lim, H.~N., Shraiman, B.~I., van
Oudenaarden, A.,
  2004. Multistability in the lactose utilization network of
  \textit{Escherichia coli}. Nature 427, 737--740.

\bibitem[{Ptashne(1992)}]{ptashne1}
Ptashne, M., 1992. A Genetic Switch: Phage $\lambda$ and Higher
Organisms, 2nd
  Edition. Cell Press \& Blackwell Scientific Publications, Cambridge, MA.

\bibitem[{Ptashne and Gann(2002)}]{Ptashne2}
Ptashne, M., Gann, A., 2002. Genes \& Signals. Cold Spring Harbor
Laboratory
  Press, Cold Spring Harbor, New York.

\bibitem[{Vilar and Leibler(2003)}]{Vilar2003}
Vilar, J. M.~G., Leibler, S., Aug 2003. {DNA} looping and physical
constraints
  on transcription regulation. J Mol Biol 331~(5), 981--989.

\bibitem[{Yagil and Yagil(1971)}]{yagil71}
Yagil, G., Yagil, E., 1971. On the relation between effector
concentration and
  the rate of induced enzyme synthesis. Biophys. J. 11, 11--17.

\end{thebibliography}

\end{document}